\documentclass[twocolumn]{aastex631}

\usepackage{amsmath}
\usepackage{gensymb}
\usepackage{enumitem}
\usepackage[acronym]{glossaries}
\usepackage{enumitem}
\usepackage{xcolor} 
\usepackage{soul}   
\usepackage{hyperref}

\usepackage{fontawesome5} 

\newcommand{\mcal}[1]{\mathcal{#1}}
\newcommand{\mrm}[1]{\mathrm{#1}}
\newcommand{\mbf}[1]{\mathrm{#1}}
\newcommand{\mbs}[1]{\boldsymbol{#1}}

\makeglossaries
\glsdisablehyper

\newacronym{mle}{MLE}{Maximum Likelihood Estimator}
\newacronym{mls}{MLS}{Maximum Likelihood Solution}
\defcitealias{Nibauer+:2023:Constraining}{N23}
\defcitealias{Law+:2009:EvidenceTriaxialMilky}{LM09}

\NewDocumentCommand{\package}{m o}{%
  \texttt{#1}\;%
  \IfNoValueTF{#2}{}{\href{#2}{\faCode}}%
}
\newcommand{\Potamides}{\package{potamides}[https://potamides.readthedocs.io]}

\shorttitle{Halo Shapes from stellar stream tracks}
\shortauthors{Wu et al.}

\begin{document}

\title{\texttt{\large Potamides}: Mapping Dark Matter Halo Shapes   from Stellar Stream Tracks in the Local Universe}

\author[0009-0003-4675-3622]{Sirui Wu}
\affiliation{DARK, Niels Bohr Institute, University of Copenhagen, Jagtvej 155A, 2300 Copenhagen, Denmark}

\author[0000-0003-3954-3291]{Nathaniel Starkman}
\affiliation{Department of Physics, Massachusetts Institute of Technology, Cambridge, MA 02139, USA}
\affiliation{Department of Astronomy, Case Western Reserve University, Cleveland, OH 44106, USA}

\author[0000-0003-0256-5446]{Sarah Pearson}
\affiliation{DARK, Niels Bohr Institute, University of Copenhagen, Jagtvej 155A, 2300 Copenhagen, Denmark}
\affiliation{DTU Space, Technical University of Denmark, Elektrovej 327, DK- 2800 Kgs. Lyngby, Denmark}

\author[0000-0001-8042-5794]{Jacob Nibauer}
\affiliation{Department of Astrophysical Sciences, Princeton University, Princeton, NJ 08544, USA}

\author[0000-0003-1808-1753]{Juan Miro-Carretero}
\affiliation{Departamento de Física de la Tierra y Astrofísica, Universidad Complutense de Madrid, Plaza de las Ciencias 2, E-28040 Madrid, Spain}
\affiliation{Leiden Observatory, Leiden University, P.O. Box 9513, 2300 RA Leiden, The Netherlands} 

\author[0009-0004-1061-7802]{David Mart\'inez-Delgado}
\affiliation{Centro de Estudios de F\'isica del Cosmos de Arag\'on (CEFCA), Unidad Asociada al CSIC, Plaza San Juan 1, 44001 Teruel, Spain}
\affiliation{ARAID Foundation, Avda. de Ranillas, 1-D, E-50018 Zaragoza, Spain}
 
\begin{abstract}\label{sec:abstract}

    Stellar streams trace the gravitational potential of their host galaxies and offer a direct probe of dark matter halo geometry. %
    Cosmological simulations predict that halo shapes depend on both baryonic physics and the nature of dark matter, yet observational constraints on halo flattening and orientation remain limited, especially for individual galaxies. %
    We present \Potamides, which utilizes the curvature of extragalactic stellar streams to derive constraints on halo shapes. %
    We apply \Potamides~ to 15 stellar streams from the Stellar Stream Legacy Survey to infer the projected axis ratios and orientation of their host halos. %
    We find that some streams in our sample exclude large regions of halo flattenings and halo orientations. %
    Systems with edge-on wrapping loops or sharp turning points yield the strongest constraints, whereas great circle-like streams remain largely uninformative. %
    All streams in our sample support a spherical halo for a given flattening direction. %
    These results demonstrate that stream morphology can provide halo shape constraints for individual external galaxies. %
    With upcoming surveys (such as Euclid, Rubin, Roman, and ARRAKIHS) expected to discover large numbers of stellar streams, this curvature-based technique will enable rapid statistical tests of dark matter and baryonic physics through the shapes and alignments of halos and disks across cosmic time. %

\end{abstract}


\section{Introduction}\label{sec:intro}

    Astrophysical observations across a wide range of scales provide compelling evidence for the existence of dark matter \citep[e.g.,][]{Zwicky:1933, Mandelbaum+:2006:GalaxyHaloMasses, Sofue:2017:RotationMassMilky, Planck+:2020:Planck2018Results}. %
    In the standard $\Lambda$ cold dark matter ($\Lambda$CDM) cosmological paradigm, galaxies form and evolve within dark matter (DM) halos, and the distribution and clumping of dark matter within these halos reflect both the underlying dark matter physics and the impact of anisotropic hierarchical assembly in the cosmic web \citep[e.g.,][]{Allgood+:2006:ShapeDarkMatter, Zhang+:2013:AlignmentsGalaxiesCosmic, Diemer+Kravtsov:2014:DependenceOuterDensity, Bullock+Boylan-Kolchin:2017:SmallScaleChallengesLCDM}. %
    In particular, the three-dimensional geometry of dark matter halos, whether close to spherical, axisymmetric (oblate, prolate), or fully triaxial, provides a diagnostic of the dark sector and of baryonic processes \citep{Kazantzidis+:2010:SphericalizationDarkMatter,Giocoli+:2026:AIDATNGProject3D}. %

    Halo shapes are commonly summarized by the principal-axis ratios $a\ge b\ge c$ of an ellipsoidal approximation to the mass or potential distribution. %
    A perfectly spherical halo has $a=b=c$, while axisymmetric configurations are classified as oblate ($a=b>c$) or prolate ($a>b=c$); when all three axes differ the halo is triaxial. %
    Collisionless $\Lambda$CDM only simulations typically predict triaxial halos, with outer regions that can be mildly prolate as a result of anisotropic accretion along filaments \citep{Jing+Suto:2002:TriaxialModelingHalo}. %
    Dark matter only simulations also predict that halos are more triaxial towards the central regions, and that more massive halos at larger redshifts are more triaxial \citep{Allgood+:2006:ShapeDarkMatter, Despali+:2017:LookHaloesCharacterization}. 
    When baryons are included, dark matter halos become more spherical towards the center, linked to the mass of the baryonic component \citep{Kazantzidis+:2010:SphericalizationDarkMatter}, and hydrodynamical simulations predict more alignment between the disk and inner dark matter halo \citep{Chua+:2019:ShapeDarkMatter, Prada+:2019:Auriga,Giocoli+:2026:AIDATNGProject3D}. %
    Alternative dark matter simulations predict differences in halo shapes, even in the case of hydrodynamical simulations. %
    Self-interacting dark matter simulations, where momentum transfer between dark matter particles occurs, produce rounder and more isotropic inner halos \citep[e.g.,][]{Dave+:2001:HaloPropertiesCosmological,Despali+:2022:ConstrainingSIDMHalo,Giocoli+:2026:AIDATNGProject3D}, while warm dark matter models, where dark matter has a higher free-streaming velocity \citep{Lovell+:2014:PropertiesWarmDark}, produce less concentrated, slightly more spherical halos than hydrodynamical $\Lambda$CDM simulations \citep[e.g.,][]{Giocoli+:2026:AIDATNGProject3D}. %
    Mapping halo shapes across different mass bins, redshifts, and distance from host halo centers could test predictions from $\Lambda$CDM and alternative dark matter models \citep[e.g.,][]{Despali+:2025:IntroducingAIDATNGProject,Despali+:2025:AIDATNGProjectDark,Giocoli+:2026:AIDATNGProject3D}. %
    
    Observationally, however, recovering halo geometry is challenging. %
    While theoretical predictions describe the full three-dimensional geometry, observational constraints for external galaxies are overwhelmingly restricted to two-dimensional projections. %
    A variety of techniques have been developed to constrain halo shapes. %
    Gravitational probes such as strong lensing can help map the mass distribution in galaxies \citep[see][for a recent review]{Shajib+:2024:StrongLensingGalaxies}, weak lensing can infer ensemble-averaged halo ellipticities from the coherent distortion of background galaxies \citep{Umetsu+:2011:ClusterMassProfiles, vanUitert+:2012:ConstraintsShapesGalaxy, Georgiou+:2021:HaloShapesConstrained}, while dynamical tracers such as satellite galaxies, globular clusters \citep{Versic+:2024:ShapesDarkMatter}, and modeling of HI gas \citep{OBrien+:2010:DarkMatterHalo} provide constraints given sufficiently precise kinematics. %
    But most techniques are limited to galaxy groups or clusters \citep[e.g., from X-rays: ][]{Reiprich+:2013:OutskirtsGalaxyClusters}. %
    For external galaxies, obtaining large samples of accurate line-of-sight velocities is often observationally expensive, and lensing-based constraints typically rely on statistical stacking and are biased towards massive systems. %

    Stellar streams offer a complementary route. %
    Streams form through the tidal disruption of dwarf galaxies or globular clusters and trace the gravitational field of their host halos \citep{Johnston+:1999:TidalStreamsProbes}. %
    In the Milky Way, several streams have been used to constrain the shape of the dark matter halo, often benefitting from 6D phase space information of individual streams (e.g.,  %
    {Sagittarius} \citep{Law+Majewski:2010:SagittariusDwarfGalaxy, Vera-Ciro+Helmi:2013:ConstraintsShapeMilky}, Palomar 5 \citep{Pearson+:2015:TidalStreamMorphology,Kupper+:2015:GlobularClusterStreams}, GD-1 (\citealt{Koposov+:2010:ConstrainingMilkyWay,Malhan+Ibata:2019:ConstrainingMilkyWay,Nibauer+Bonaca:2025:Tilted}), or using multiple streams (\citealt{Lux+:2013:ConstrainingMilkyWay,Bovy+:2016:ShapeInnerMilky,Bonaca+Hogg:2018:InformationContentCold,Bariego-QuintanaLlanes-Estrada:2024:TorsionStellarStreams}). %
    In external galaxies, the streams can similarly constrain individual galaxy halos, in projection. Likewise, if multiple streams are detected around one host halo, one could constrain that halo at different galactocentric radii, since streams are sensitive to the local acceleration field \citep{Bonaca+Hogg:2018:InformationContentCold, Nibauer+:2022:ChartingGalacticAccelerations}. %

    For external galaxies, ongoing and upcoming surveys will uncover unprecedented numbers of streams for which two-dimensional morphology is often the most readily available information (e.g., {\it Euclid} \citep{Racca+:2016:EuclidMissionDesign}, The Rubin Observatory \citep{LSST+:2019:LSSTScienceDrivers}, The Nancy Grace Roman Space Telescope ({\it Roman}; \citealt{Spergel+:2015:WideFieldInfrarRedSurvey}), and ARRAKIHS  (\citealt{Guzman+:2022:DUNESARRAKHISSpace})). %
    This opens up the potential to use streams to map halo shapes across various mass and redshift ranges. %
    As our goal is to test dark matter predictions with streams, we need rapid techniques to infer halos shapes across large samples. %
    Several groups have developed new stream modeling techniques to leverage extragalactic streams to learn about dark matter halos \citep[e.g.,][]{Fardal+:2013:InferringAndromedaGalaxys,Foster2014,Amorisco+:2015:DwarfGalaxysTransformation,Pearson+:2022:MappingDarkMatter}, some of which use morphologies of the streams alone \citep{Nibauer+:2023:Constraining, Walder+:2025:ProbingDarkMatter, Nibauer+Pearson:2025:TestingDarkMatter, Chemaly+:2026:HierarchicalBayesianInference, Starkman+:2026:EuclidStreamsPilot}. %

    The curvature-based method developed by \citet[][hereafter N23]{Nibauer+:2023:Constraining} is compelling because it relies solely on the morphology of 2D projected stream tracks. %
    This method is motivated by a dynamical assumption known as the stream ``coherence condition'': along a stellar stream, the local 2D curvature vector of the projected track should point within $90^{\circ}$ of the local gravitational acceleration vector projected onto the plane of the sky. %
    This geometric alignment must be satisfied at the true line-of-sight distance of the stream segment; otherwise, the ensemble of orbits making up the stream would not have curved in the observed direction in projection. %
    This generalized $90^{\circ}$ condition cleanly rules out potential models that are fundamentally incompatible with the stream's observed 2D curvature. %
    \citetalias{Nibauer+:2023:Constraining} demonstrated that streams exhibiting sharp changes in concavity or tight wrapping loops are particularly constraining, as they sample rapidly changing directions of the local acceleration field. %
    Applying this technique to the stellar stream around NGC 5907 \citep{Martinez-Delgado+:2008:GhostDwarfGalaxy, vanDokkum+:2019:DragonflyImagingGalaxy, Muller+:2019:HuntingGhostsIconic}, \citetalias{Nibauer+:2023:Constraining} inferred a halo flattened perpendicularly to the stellar disk, showcasing the method's ability to probe dark matter halo-baryonic alignments. %
  
    In this work, we present the new curvature-based code, \Potamides~\footnote{\url{https://github.com/xggs-dev/potamides}} \citep{Potamides:2026}, developed from the core ideas presented in \citetalias{Nibauer+:2023:Constraining}, and extended to more rigorous treatment of hyperparameters. %
    We apply \Potamides~ to 15 stellar streams from the Stellar Stream Legacy Survey \citep{Martinez-Delgado+:2023:SSLS, Miro-Carretero+:2024:SSLS}, which is a low redshift ($z<0.02$) sample of streams from the DESI Legacy Imaging Surveys (DESI LS: \citealt{DESI+:2019}) and from DECam data from the Dark Energy Survey (DES) \citep{DES+:2018:DR1}. %
    We use \Potamides~ to infer the projected halo geometries using only the stream ridgelines traced in the plane of the sky. %
    While most of the streams in our sample are relatively close to their host galaxy (within 3 scale lengths of the disks), future applications of the code to other datasets could reveal halo shapes at larger galactocentric radii and halo shape trends with redshift (e.g., \citealt{Starkman+:2026:EuclidStreamsPilot}). %

    This paper is organized as follows. %
    In \autoref{sec:data} we describe the stellar stream sample. %
    In \autoref{sec:methods} we describe our methods using an on-sky tangent-plane coordinate system, our stellar stream track extraction, the potential model used, and our inference pipeline. %
    We present the results of our analysis in \autoref{sec:results} and discuss our findings in \autoref{sec:discussion}. %
    We summarize our conclusions in \autoref{sec:conclusion}. %

\section{Data} \label{sec:data}
    \begin{figure*}
        \centering
        \includegraphics[width=\linewidth]{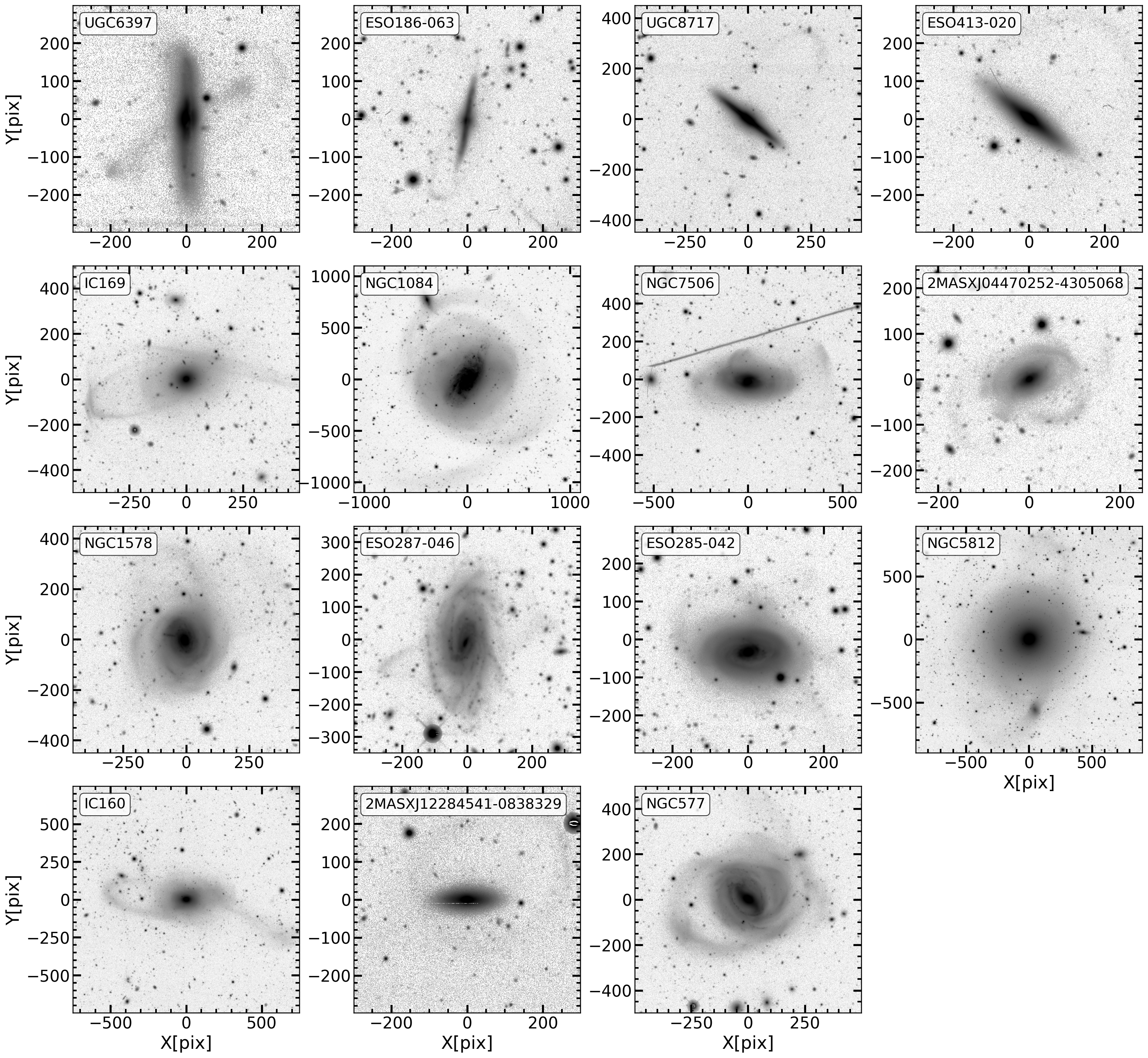}
        \caption{
            R-band image of 15 sample galaxies from SSLS. %
            The name of each galaxy is indicated in the label at the top-left corner of each panel. %
            The X and Y axis are given in the unit of pixels. %
            Ticks label are displayed only along the left and bottom. %
            The field of view varies between panels, as reflected by the different axis ranges. %
        }
        \label{fig:data:all_galaxies}
    \end{figure*}
    We selected our galaxy sample from the Stellar Stream Legacy Survey (SSLS; \citealt{Martinez-Delgado+:2023:SSLS, Miro-Carretero+:2024:SSLS}, hereafter SSLS Paper~I and Paper~II). %
    This survey systematically searches for low-surface-brightness tidal features associated with dwarf-galaxy accretion around nearby galaxies ($z\lesssim 0.02$). %
    The SSLS utilizes a custom re-reduction of public imaging from the DESI Legacy Imaging Surveys (DESI LS) \citep{DESI+:2019}, incorporating reprocessed DECam data from the Dark Energy Survey (DES) \citep{DES+:2018:DR1}. %
    The SSLS parent sample relies on the HyperLeda database for recessional velocities and luminosities. %
    Foreground-extinction-corrected $K$-band magnitudes are primarily sourced from the 2MASS survey \citep{Skrutskie+:2006:TwoMicronAll, Schlafly+Finkbeiner:2011:MeasuringReddeningSloan}. %

    The survey selects targets based on four primary criteria: %
    \begin{enumerate}[leftmargin=*]
        \item \textbf{Stellar mass limit:} A $K$-band luminosity cut of $M_K < -19.6$~mag, ensuring a minimum stellar mass of roughly $10^9 M_\odot$. %
        \item \textbf{Distance window:} A Local Group rest-frame velocity between 2000 and $7000~\mathrm{km~s^{-1}}$, corresponding to distances of approximately 30 to 100~Mpc. %
        The lower bound guarantees that a standard $30'\times30'$ field of view covers a typical virial radius (250 to 300~kpc), while the upper bound ensures sufficient image resolution for stream detection. %
        \item \textbf{Galactic latitude:} An absolute Galactic latitude of $|b|>20^\circ$ to minimize dust extinction and stellar crowding near the Galactic plane. %
        \item \textbf{Environmental isolation:} Any neighboring galaxy within a 1~Mpc projected radius and a velocity difference of $|\Delta V|<250~\mathrm{km~s^{-1}}$ must be at least 2.5 magnitudes fainter than the host in the $K$ band. %
    \end{enumerate}

    These criteria yielded a parent sample of approximately 3100 targets. %
    This compilation includes roughly 940 Milky Way analogues, defined by their $K$-band luminosities ($-24.6 < M_K < -23.0$) and local environment \citep{Geha+:2017:SAGASurveySatellite}. %
    SSLS Paper~I established the proof of concept by visually identifying 24 new stellar streams. %
    Subsequently, SSLS Paper~II conducted a systematic visual search across the DES footprint, cataloging 63 streams (including 58 newly reported systems) around approximately 700 nearby galaxies. %
    
    For the present work, we analyzed a pilot sample of 15 host galaxies drawn from the combined SSLS compilation. %
    Rather than forming a statistically complete subset, this working group serves to demonstrate our curvature-based inference method. %
    We successfully applied our ridge-line tracing and curvature measurement procedures (\autoref{sec:methods:ridgeline_tracing}) to all 15 systems. %

\section{Methods}\label{sec:methods}

    In this section, we describe the components of our geometric inference pipeline. %
    First, we define the sky-plane coordinate system used throughout (\autoref{fig:coordinate}). %
    We then explain how we extract a two-dimensional stream ridgeline (``track'') from imaging data and represent it as a smooth twice-continuous ($C^2$) curve, yielding well-defined tangent and curvature vectors (\autoref{sec:methods:ridgeline_tracing}). %
    Next, we specify the halo gravitational potentials used to generate the sky-plane acceleration field and define the two projected halo parameters we constrain: the isopotential axis ratio $q$ (flattening) and the major-axis orientation $\theta$ (orientation) (\autoref{sec:methods:potential}). %
    Finally, we construct a likelihood based on a sign-only alignment test between the track's curvature direction and the halo's in-plane acceleration direction, and we describe how likelihood contributions from multiple stream segments are combined (\autoref{eq:methods:likelihood}). %

    \subsection{Coordinate System and Physical Scale}\label{sec:methods:coordinates}
    \begin{figure}
            \centering
            \includegraphics[width=\linewidth]{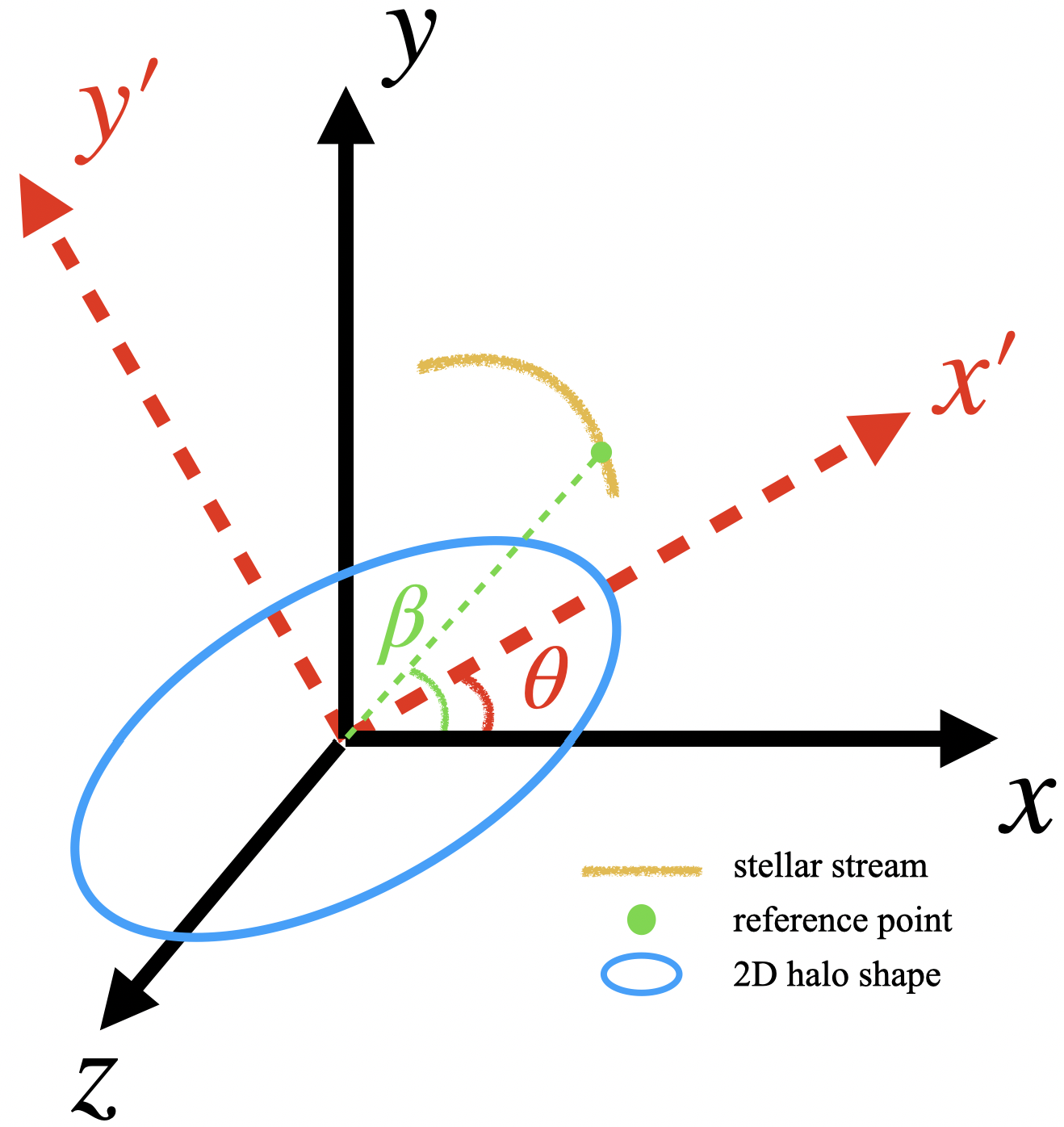}
            \caption{%
                Schematic of the coordinate system. %
                The origin is at the host center; the $x$--$y$ plane coincide with the sky plane, and the $z$-axis points toward the observer. %
                The $x'$-$y'$ axes are aligned with the principal axes of the projected halo (blue ellipse). %
                $\theta$ is the orientation of the projected halo major axis, and $\beta$ is the position azimuth of a reference point on the stream (gold curve); both are measured counterclockwise from the $x$-axis. %
            }
            \label{fig:coordinate}
        \end{figure}
        
        We perform all calculations using the flat-sky approximation in a local tangent-plane coordinate system. %
        The line of sight defines the $z$-axis. %
        We define Cartesian coordinates $(x,y)$, centered on the host galaxy (\autoref{fig:coordinate}), where the positive $y$ axis points South and the positive $x$ axis points West, placing East to the left. %
        Because our targets subtend only a few arcminutes, this approximation introduces negligible error.  %
        Any residual rotation relative to the detector is absorbed into the inferred halo major-axis orientation $\theta$. %
        To convert angular projected separations into physical distances (kpc),  we adopt {\it Planck} 2018 cosmological parameters \citep{Planck+:2020:Planck2018Results} as implemented in \texttt{astropy} \citep{Astropy+:2022}. %

    \subsection{Stream Ridgeline Tracing} \label{sec:methods:ridgeline_tracing}

        Stellar streams are low--surface-brightness features \citep{Bullock+Johnston:2005:TracingGalaxyFormation,Crnojevic+:2016:ExtendedHaloCentaurus,Martinez-Delgado+:2023:SSLS}, and for extragalactic systems their detection typically relies on careful background modelling and/or visual contrast enhancement \citep{Pearson+:2022:HoughStreamSpotter,Sola+:2025:STRRINGS,Benitez-Walz+:2026:ContrastiveLearningExtragalactic}. %
        Crucially, our analysis requires only the \emph{ridgeline} (the on-sky path of the stream) rather than stream widths or surface-brightness profiles. %
        This requirement significantly simplifies the characterization challenge. %
        We adopted a tracing procedure designed to prioritize smoothness, yielding a differentiable representation from which reliable tangents and curvatures can be measured. %

        For the galaxies in our sample, stream morphologies are readily identifiable through visual contrast stretching of the original images. %
        We therefore traced the streams directly on these contrast-stretched images. %
        While modelling and subtracting the host-galaxy light is a common approach to reveal faint tidal features  \citep[e.g.,][]{Kim+:2012:EarlytypeGalaxiesTidal, Mantha+:2019:StudyingPhysicalProperties, Sola+:2022:CharacterizationLowSurface, Rutherford+:2024:SAMIGalaxySurvey}, this process can introduce subtraction artifacts, such as over-subtraction rings or dipolar residuals. 
        These systematics can artificially mimic or distort the geometry of low-surface-brightness structures. %
        By bypassing this step, we prioritized geometric fidelity for the specific systems analyzed in this work. %
        
        For each system, we constructed an ordered set of control points $\{\boldsymbol{x}_i\}$ along the apparent ridgeline using an interactive annotation tool. %
        Annotators traced the visually continuous path of the stream on contrast-stretched images, and we fitted the resulting point set with a smooth spline. %
        Formally, we represented this track as a continuous planar curve $\vec{x}(\gamma)=(x(\gamma),y(\gamma))$ with parameter $\gamma\in[-1,1]$. %
        In practice, we reparameterized $\gamma$ to be proportional to the cumulative arc length $s$, ensuring that uniform sampling in parameter space corresponds to uniform physical spacing along the stream. %

        Because our likelihood analysis relies on comparing the track's bending direction to the halo's unit vector gravitational acceleration in the sky-plane, $\hat{\mathbf{a}}_{xy}$, the fitted track must be $C^2$. %
        This regularity ensures that the geometric curvature vector, defined as the second derivative with respect to arc length ($\boldsymbol{\kappa} \equiv d^2\vec{x}/ds^2$), and its unit direction $\hat{\boldsymbol{\kappa}}$ are well defined and numerically stable. %
        The vector $\boldsymbol{\kappa}$ represents the intrinsic geometric curvature, which is distinct from the raw second derivative with respect to an arbitrary spline parameter. %
        We provide details of the regularization applied to suppress spurious small-scale concavity flips are provided in \autoref{sec:appendix:optimizing_the_track}. %

        We perform the tracing in the $r$ band. %
        For the systems in our sample, the stream morphology is consistent across $g/r/z$ images. %
        The $r$ band typically offers a favorable balance between depth and background stability, providing a high-contrast stream appearance under our stretching procedure. %

        During the preparation of this work, the STRRINGS catalog \citep{Sola+:2025:STRRINGS} presented a complementary sample of long, narrow streams traced on model-subtracted residual images constructed for SGA2020 galaxies \citep{Moustakas+:2023:SienaGalaxyAtlas}. %
        Because visual contrast stretching provides sufficient clarity for robust ridgeline annotation in our current sample, we did not employ model subtraction, thereby avoiding the potential geometric artifacts associated with such techniques. %
        However, utilizing carefully constructed residual products like those in STRRINGS will be crucial for extending our curvature-based method to streams that are more severely obscured by their host galaxies. %

    \subsection{Gravitational Potential}\label{sec:methods:potential}

        We modeled the host dark-matter halo using a generic three-dimensional logarithmic potential. %
        Unlike triaxial NFW profiles, which require computationally expensive numerical integrals to evaluate the acceleration field, logarithmic potentials yield fully analytic gradients. %
        Furthermore, they naturally produce the flat rotation curves typical of the galactocentric radii where our target streams reside. %
        A general triaxial logarithmic potential can be expressed using a quadratic form: %
        \begin{equation}\label{eq:methods:potential:general}
            \Phi_{\rm general}(\mathbf{x}) = v_{\rm halo}^2\,\ln\!\left(r_{\rm halo}^2 + \sum_{i,j \in \{x,y,z\}} c_{ij} x_i x_j \right),
        \end{equation}
        where $v_{\rm halo}$ is a velocity scale, $r_{\rm halo}$ is a core radius, and $c_{ij}$ are the elements of a $3 \times 3$ symmetric matrix governing the shape and orientation of the equipotential surfaces. %
        The corresponding 3D acceleration field is explicit and analytic: %
        \begin{equation}\label{eq:methods:potential:acc_general}
            a_k = -\frac{\partial \Phi}{\partial x_k} = - \frac{2 v_{\rm halo}^2}{r_{\rm halo}^2 + \sum_{i,j} c_{ij} x_i x_j} \sum_{j} c_{kj} x_j .
        \end{equation}
        Because our inference relies exclusively on the \emph{direction} of the acceleration field, the scalar prefactor in \autoref{eq:methods:potential:acc_general} drops out upon normalization to a unit vector. %
        Consequently, the 3D acceleration direction is governed entirely by the symmetric matrix $c_{ij}$. %
        This matrix contains six independent elements. %
        Factoring out an overall geometric scale, which does not alter the direction vector, leaves five independent degrees of freedom required to specify the 3D halo orientation and shape. %
        Our inference relies exclusively on the \emph{direction} of the projected acceleration field, so the scalar prefactor in \autoref{eq:methods:potential:acc_general} drops out upon normalization to a unit vector. %
        The 3D acceleration direction is therefore governed entirely by the symmetric matrix $c_{ij}$, which contains six independent elements; factoring out an overall geometric scale leaves five degrees of freedom required to specify the 3D halo orientation and shape. %
        Since we extract constraints from the 2D morphology of stellar streams, we evaluate the acceleration field strictly on the sky plane. %
        Setting $z=0$ reduces the spatial quadratic form to a $2\times2$ symmetric matrix with three independent elements ($c_{xx}, c_{yy}, c_{xy}$), and normalizing out the overall 2D scale leaves exactly two independent parameters. %
        We parameterize these two degrees of freedom using the projected minor-to-major axis ratio (the flattening) $q \in (0, 1]$ and the projected major-axis orientation $\theta \in [-90^{\circ}, 90^{\circ}]$, measured counterclockwise from the $x$-axis (\autoref{fig:coordinate}). %
        These parameters describe equipotential contours rather than isodensity contours; while a flattening below $q \sim 0.7$ would be extreme in density, it still provides a useful description of the perpendicular acceleration field. %
        We note that \citetalias{Nibauer+:2023:Constraining} showed near-zero-curvature segments to be particularly constraining, as their projected acceleration vectors should nearly align with the stream track. %
        In this work, we do not adopt this condition, making our results conservative but robust to misidentification of linear features. %

        In practice, we implemented this 2D projected field using the \texttt{galax} dynamics package, which provides performant implementations of all the common analytic potentials, specifically the triaxial logarithmic parametrization introduced by \citet{Law+:2009:EvidenceTriaxialMilky}. %
        This parametrization restricts the general 3D logarithmic potential by fixing one principal axis to the internal $z$-axis, parametrizing the geometry with three axis lengths and an internal twist angle. %
        To map this parametrization to our general 2D framework, we aligned the model's $z$-axis with the line of sight and disabled the internal twist. %
        We then fixed the major axis scale to unity, adopted our parameter $q$ as the in-plane minor-to-major axis ratio, and specified the arbitrary on-sky orientation using a global coordinate rotation by $\theta$. %
        Following this procedure, our identifiable halo model is cleanly specified by $\vec{m}=(q,\theta)$. %
        
    \subsection{Likelihood}\label{eq:methods:likelihood}

        Following \citetalias{Nibauer+:2023:Constraining}, our likelihood is built from a simple geometric test motivated by the stream ``coherence condition''. %
        This dynamical assumption posits that at any point along the stream track, the projected acceleration vector must not oppose the observed curvature vector. %
        If this condition is violated, the orbits of the constituent stars would rapidly decouple from the observed stream track, destroying the coherent, long-term morphology of the stream. %
        To maintain this stable geometry, the ridgeline at a given location along a stream must bend toward the same sky-plane side as the local in-plane gravitational acceleration predicted by the potential. %
        Equivalently, at most locations, the angle $\varphi$ between the \emph{planar} unit curvature vector of the track, $\hat{\boldsymbol{\kappa}}$, and the \emph{planar} unit gravitational acceleration vector, $\hat{\mathbf{a}}_{xy}$, must satisfy $\varphi<90^\circ$. %
        We discuss the orbit-level motivation for this sign condition, along with known failure modes related to ridgeline extraction and sky-plane projection, in \autoref{sec:discussion:limitations}. %
        Using the smooth parametric track $\vec{x}(\gamma)$ constructed in \autoref{sec:appendix:optimizing_the_track}, where $\gamma \in [-1,1]$ is approximately proportional to the arc length, we evaluate the unit curvature vector $\hat{\boldsymbol{\kappa}}_i$ at each reference point $i$. %
        For a trial set of projected halo parameters $\vec{m}=(q,\theta)$, we compute the corresponding unit in-plane acceleration vector $\hat{\mathbf{a}}_{xy,i}$ using \autoref{eq:methods:potential:acc_general}. %
        
        Based on the geometric assumption, we define pass($f_1$)/fail($f_2$) fractions: %
        \begin{align}\label{eq:fraction}
            f_1 &\equiv \frac{1}{N}\sum_{i=1}^N \frac{1}{2}\,\bigl|1+\operatorname{sgn}(\hat{\mathbf{a}}_{xy,i}\cdot\hat{\boldsymbol{\kappa}}_i)\bigr|
                    \;=\; \frac{n_1}{N},\\[4pt]
            f_2 &\equiv 1-f_1 \;=\; \frac{n_2}{N},
        \end{align}
        where $N$ is the total number of reference points, $n_1$ is the number with $\varphi_i<90^\circ$,
        and $n_2=N-n_1$ is the number with $\varphi_i\ge 90^\circ$. %
        
        We then write the (unnormalized) log-likelihood as %
        \begin{equation}\label{eq:likelihood}
            \ln \mathcal{L}(\{\hat{\kappa}_i\}\mid \vec{m})=
            \begin{cases}
                N\bigl(f_1\ln f_1 + f_2\ln f_2\bigr), & n_2<50\%,\\[3pt]
                -\infty, & n_2\ge 50\%,
            \end{cases}
        \end{equation}%
        which enforces our baseline geometric assumption by assigning $-\infty$ when a majority of points violate $\varphi<90^\circ$. %
        From \autoref{eq:likelihood}, ensuring that $n_1/N\to 1$ we have $\ln \mathcal{L}\to 0$. %
        The entropy-like term $N(f_1\ln f_1+f_2\ln f_2)$ is symmetric in $(f_1,f_2)$ and attains its %
        minimum at $f_1=f_2=1/2$ (i.e., $n_2=50\%)$. %
        Because we use an \gls{mle}-style comparison, the relevant quantity is the normalized likelihood %
        $\mathcal{L}/\mathcal{L}_{\max}$, which for a single stream depends only on the fraction $n_1/N$ %
        and is therefore independent of the absolute sample size $N$. %

        \subsubsection{Combining multiple segments}\label{sec:methods:likelihood:combining}

            Some systems contain several stellar streams, or a single stream  divided into
            segments due to masking or surface-brightness limits.
            For each segment $s$, we compute the profile log-likelihood $\ln\mathcal{L}_s$ from \autoref{eq:likelihood} using the corresponding counts $(n_{1,s}, n_{2,s}, N_s)$.
            To prevent densely sampled segments from dominating the joint constraint, we combine the segment likelihoods using weights:
            \begin{equation}\label{eq:seg_comb}
                \ln \mathcal{L}_{\rm SC}(\vec{m})
                = \sum_{s=1}^{S} w_s\, \ln \mathcal{L}_s(\vec{m}),
            \end{equation}
            where the weights $w_s$ satisfy $\sum_s w_s = 1$.
            By default, we adopt arc-length--based weights so that each unit of stream length contributes comparably to the total constraint. Concretely we use:
            \begin{equation}
                w_s = \frac{(L_s/N_s)}{\sum_{j=1}^{S} (L_j/N_j)},    
            \end{equation}
            where $L_s$ is the arc length of segment $s$ and $N_s$ is its number of reference points. %
            This normalization avoids giving undue leverage to segments sampled at higher point density. %

\section{Results}\label{sec:results}

    In this section, we present representative examples illustrating how individual stellar streams constrain the projected halo flattening $q$ and orientation $\theta$. %
    We begin with systems yielding tight constraints, including hosts with and without stellar disks, and cases where we combine likelihoods from multiple streams around a single galaxy.
    Next, we discuss ``turning-point'' streams (\autoref{sec:results:galaxy_Ushape}), characterized by a single direction change about a point. %
    This is a geometric effect and does not necessarily correspond to an orbital turning point. %
    Finally, we consider cases providing only weak constraints, mostly notably nearly circular or ring-like streams, where the curvature-based method proves less informative (\autoref{sec:results:galaxy_weak}). %
    The Appendix summarizes the remaining systems (\autoref{app:other_results}). %

\subsection{Well-Constrained Galaxies}\label{sec:results:galaxy_good}

\begin{figure*}[htbp]
\centering   
\includegraphics[width=\linewidth]{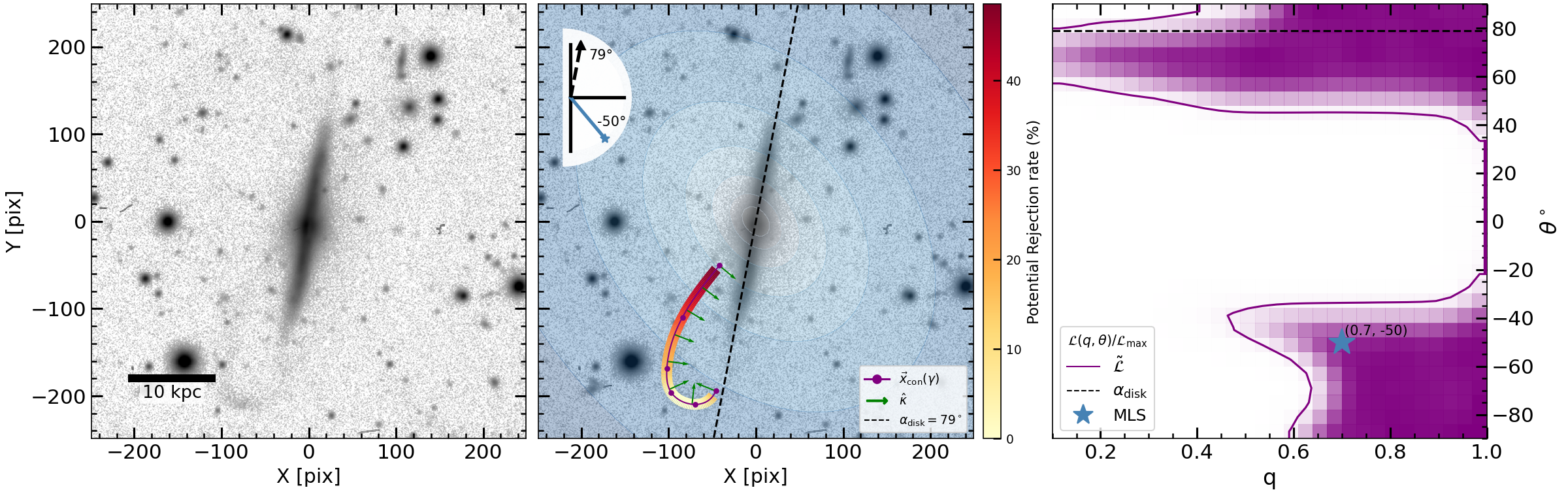}
\caption{
ESO 186-063: Likelihood inference from a single stellar stream. %
Left: Raw $r$-band image with a 10 kpc scale bar. %
Middle: The fitted stream track $\vec{x}_{\rm con}(\gamma)$ (purple curve) with control points (purple nodes) and local unit curvature vectors $\hat{\kappa}$ (green arrows), overlaid on the halo equipotentials (light blue contours) of the MLS marked by the star in the right panel. %
The track colour encodes the local potential-rejection rate (redder $=$ stronger constraint; the outermost 5\% at each end is excluded); the colour coding is shown only in this figure. %
The dashed line and unit-circle inset show the disk ($\alpha_{\rm disk}=79\degree$) and MLS halo projected major-axis orientation ($\theta=-50\degree$). %
Right: Normalized profile likelihood $\mathcal{L}(q,\theta)/\mathcal{L}_{\rm max}$ over halo projected flattening $q$ and orientation $\theta$. %
White regions are excluded by the curvature constraint. %
Purple shading indicates the feasible region (darker = higher likelihood); %
the darkest plateau represents equally likely maximum-likelihood models, among which the star marks one representative MLS at $(q,\theta)=(0.7,-50\degree)$.
The dashed line marks the disk orientation. %
}
\label{fig:ESO186-063}
\end{figure*}    
            
\autoref{fig:ESO186-063} shows the curvature-based inference for the stellar stream around ESO~186--063, which provides one of the tightest constraints in our sample. %
At this viewing angle, the stream appears to follow a highly radial trajectory at small galactocentric distance and shows a sharp turn at larger distances. %
We only annotate the robustly visbile regions (see middle panel), but it is possible that an extension of the stream exists towards the lower right, potentially giving the stream a more symmetric U-shape morphology. %
The color coding in the middle panel shows where our method is most constraining. Darker colors means that more potentials are rejected due to the geometric alignment test, where lighter colors show less constraining parts of the stream. %
The likelihood map in the $(q,\theta)$ plane (right panel) is highly anisotropic. %
The purple-shaded region shows the allowed parameter space, bounded by the solid purple contour of the 95\% confidence region. %
The darkest purple region represents the maximum-likelihood plateau, where 
all models share the same likelihood value. %
White regions correspond to halo geometries excluded by the geometric alignment test. %
The blue star marks one example of a \gls{mls}, $(q, \theta) = (0.7, -50\degree)$, which we adopt as the fiducial model for halo equipotential contours shown in light blue in the middle panel. %
The black dashed line indicates the orientation of the disk major axis. %
The right panel demonstrates that spherical halos ($q=1$) are allowed at all orientations: the stream morphology of ESO~186--063 does not generate sufficient geometric tension to exclude this case. %
Halo orientations in the range $-30\degree \lesssim \theta \lesssim 50\degree$ are ruled out for nearly all flattenings except the spherical case. %

To illustrate why this range is ruled out, consider a strongly flattened halo with $(q, \theta) = (0.3, -65\degree)$ and a reference point near the lower right end of the stream, where the curvature direction points towards the upper left. %
If the potential is strongly flattened (e.g., $q=0.3$), the acceleration would have a strong component along the short axis of the halo, and would point roughly toward the upper right, producing an angle $\varphi > 90\degree$,
which is ruled out from the geometric alignment test. %

We note that even extremely flattened halos are allowed when $\theta$ lies close to the disk orientation, $\alpha_{\rm disk} = 79\degree$, which we discuss further in \autoref{subsec:disk}. %
ESO~186--063 provides a clear example of how in projection a radially plunging, tightly curved stream can yield strong, orientation-dependent constraints on the host halo geometry. %

        \begin{figure*}
            \centering
            \includegraphics[width=\linewidth]{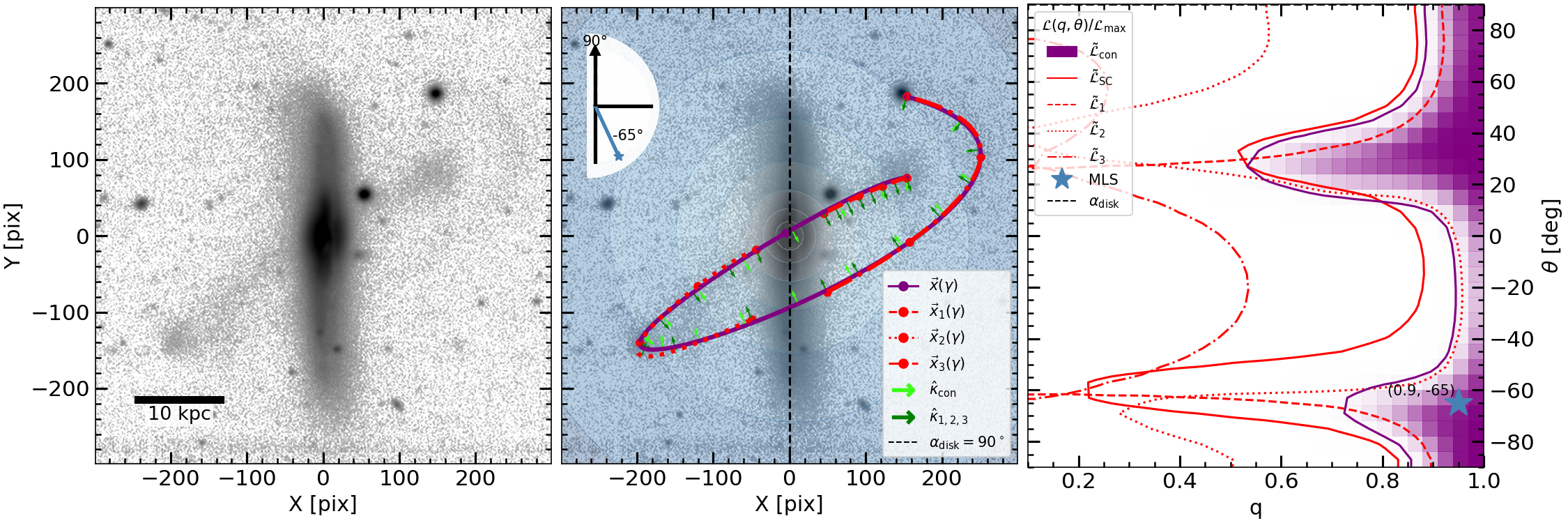}
            \caption{%
                UGC6397: Likelihood inference from a visually interrupted stream. %
                Left: Raw $r$-band image with a 10 kpc scale bar. %
                Middle: Comparison of two track reconstructions strategies for the visually interrupted stream, overlaid on the halo equipotentials (light blue contours) of the MLS marked by the star in the right panel. %
                The continuous track ($\vec{x}_{\rm con}(\gamma)$, purple) and per-segment tracks ($\vec{x}_{1,2,3}(\gamma)$, red dashed/dotted/dash-dotted) share the same control points (red nodes); curvature vectors are shown in light green ($\hat{\kappa}_{\rm con}$, continuous) and dark green ($\hat{\kappa}_{1,2,3}$, segments). %
                The dashed line and unit-circle inset show the disk ($\alpha_{\rm disk}=90\degree$) and MLS halo major-axis orientation ($\theta=-65\degree$). %
                Right: Normalized profile likelihood $\mathcal{L}(q,\theta)/\mathcal{L}_{\rm max}$ over halo projected flattening $q$ and orientation $\theta$. %
                Purple shading indicates the feasible region (darker = higher likelihood) for the inference from the continuous track $\vec{x}_{\rm con}(\gamma)$;
                the darkest plateau represents equally likely maximum-likelihood models, among which the star marks one representative MLS at $(q,\theta)=(0.9,-65\degree)$. %
                Solid red line: feasible boundary from the combined segment inference $\tilde{\mathcal{L}}_{\rm SC}$;
                red dashed/dotted/dash-dotted lines: per-segment feasible boundaries $\tilde{\mathcal{L}}_{1,2,3}$.
                For all red lines, the feasible region lies to their right (higher $q$), while the region to the left is excluded. %
            }
            \label{fig:UGC6397}
        \end{figure*}

        \autoref{fig:UGC6397} shows the curvature-based inference for the stream around UGC~6397, which provides strong but qualitatively different constraints. %
        The host is a disk galaxy viewed nearly edge-on, with a bright, elongated central body consistent with a prominent bulge. %
        The projected morphology of the stream appears to form a single coherent loop, wrapping around the host, with a nearly edge-on orbital plane. %
        
        Similar to the radially plunging stream in ESO~186--063, the UGC~6397 stream passes near the galaxy center, in projection. When we analyse the stream as one continuos structure, the likelihood map (right panel) shows that for most halo orientations only nearly spherical halos are allowed; only near $\theta\approx 30\degree$ are mildly flattened halo ($q\gtrsim 0.7$) permitted. %
        
        UGC~6397 provides a useful test of the robustness of the inference to track reconstruction. %
        The stream is likely physically continuous, but appears fragmented into three visible segments due to obscuration by the bright host galaxy. %
        We therefore analyse the system in two complementary ways: %
        (\emph{i}) treating the stream as three separate segments, each fit independently with per-segment likelihoods $\tilde{\mathcal{L}}_{1,2,3}$, and %
        (\emph{ii}) fitting a single continuous track $\vec{x}_{\rm con}(\gamma)$ bridging the obscured region. %
        The two approaches produce slightly different local curvature directions in localized regions (see light and dark green arrows in the middle panel) because the track is required to be twice continuously differentiable, resulting in small differences near the endpoints of each segment. %
        The slight differences does not affect the overall result. %

        The continuous fit ($\tilde{\mathcal{L}}_{\rm con}$, black line and purple region in the third panel) imposes more stringent constraints than the combination of the three segments ($\tilde{\mathcal{L}}_{\rm SC}$, red solid line), and each segment analysed individually also produce less stringent constraints (see dashed/dotted/dash-dotted lines).  %
        For $-70\degree \lesssim \theta\lesssim -50\degree$, the continuous fit excludes all but nearly spherical halos, whereas the segmented combination allows halos as flat as $q\approx 0.2$, allowing for both mildly and extremely flattened configurations. %
        
        The strong constraints in the continuous case arise from the interpolated segment connecting $\vec{x}_1(\gamma)$ and $\vec{x}_2(\gamma)$: this segment is nearly straight, so its curvature-direction span is small, yet it passes close to the galactic center where the acceleration direction rotates rapidly, producing a large mismatch between the two fields (\autoref{subsec:sensitivity}). %
        We note, however, that this result is conditional on our assumption that the host center of mass coincides with the center of light. %
        If that assumption were relaxed, the detailed constraints associated with the bridge segment would change, although not necessarily become weaker. %
        
        In practice, for streams that are plausibly continuous, we use the continuous reconstruction as the fiducial case and the segmented reconstruction as a robustness check; a more detailed discussion is given in \autoref{sec:discussion:continuity_and_length}. %

    \subsection{Turning-point Streams}\label{sec:results:galaxy_Ushape}

        We next consider a class of systems we refer to as turning-point streams. %
        These are systems that contain a single reversal in direction, producing a U-shaped morphology on one side of the host galaxy. %
        Morphologically, these systems are characterized by two more straight segments with lower curvature and a compact, higher-curvature arc. %
        Unlike the wrapped loop in UGC~6397 (\autoref{fig:UGC6397}), these systems do not enclose the host galaxy in projection. %
        ESO~186--063 \autoref{fig:ESO186-063} can also be considered as a member of this class, although only one of its lower-curvature, straight segments is clearly detected in current data. %

        \begin{figure*}
            \centering
            \includegraphics[width=\linewidth]{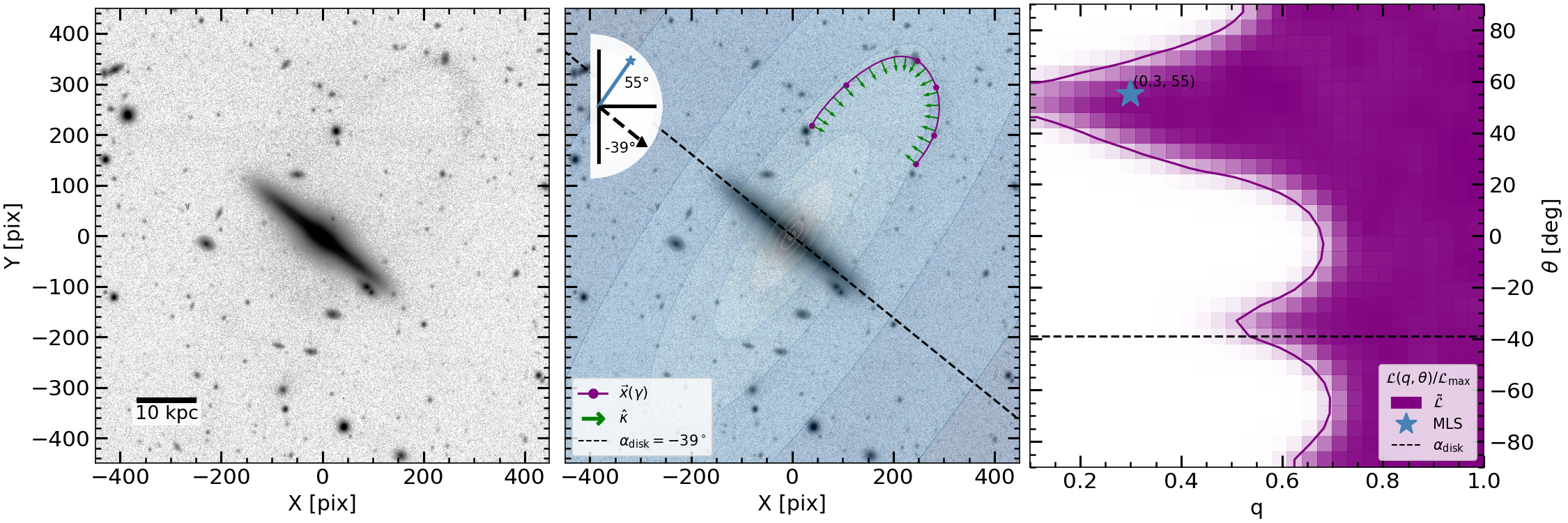}
            \caption{%
                UGC8717: Likelihood inference from a single stellar stream. %
                Colors, symbols, and panel arrangement follow \autoref{fig:ESO186-063}. %
                In this system, the disk orientation is $\alpha_{\rm disk}=-39\degree$, and the selected MLS is at $(q,\theta)=(0.3,55\degree)$. %
            }
            \label{fig:UGC8717}
        \end{figure*}

        \autoref{fig:UGC8717} shows our first example of a turning-point stream, the stellar stream around UGC~8717. %
        The host is a disk galaxy with a well-defined major axis, indicated by the dashed line. %
        The stream traces a compact arc that bends sharply through a well-defined turning point at a moderate projected galactocentric distance (see scale bar on left panel), and which is confined to one side of the galaxy. %
        On either side of the turning point, the stream is relatively straight and points roughly towards the host. %
        The curvature direction along each of these two segments are approximately constant, one pointing roughly to the lower right, and the other to the upper left. 
        Moreover, the reference points (see green vectors) on these two segments span a wide range of position azimuths (i.e., the angle $\beta$ between the vector from the galaxy center to the stream reference point and the positive $x$-axis shown in green in \autoref{fig:coordinate}), which means that the planar acceleration span is much larger than the curvature direction span. %
        This mismatch makes the lower-curvature segments in the U-shaped stream particularly sensitive to halo geometry: even a modest change in $(q, \theta)$ can flip the sign of $\hat{\boldsymbol{\kappa}}\cdot\hat{\mathbf{a}}_{xy}$ at multiple points, providing tighter constraints. %

        In \autoref{fig:UGC8717} (middle), the blue contours show the equipotential contours of a representative \gls{mls} model selected from the likelihood plateau (blue point, right panel). %
        In the example shown, the major-axis halo orientation $\theta=55\degree$ is close to the symmetry axis of the U-shaped stream, i.e. where the stream turns most sharply. In the right panel, the purple region spans all flattenings at $\theta=55\degree$ indicating a weak constraint. %
        For other halo orientations, the constraints are significantly tighter. %
        Deviations from the stream's symmetry axis create greater tension during the geometric alignment test. %
        For example, at $\theta\sim0\degree$, the allowed projected flattening is restricted to $0.7\lesssim q\leq1$. %
        More strongly flattened halos in this orientation, the acceleration directions gain a larger component along the $x$-axis, which leads to more reference points with  $\hat{\boldsymbol{\kappa}}\cdot\hat{\mathbf{a}}_{xy}<0$ along the stream ridgeline. %
        
        \begin{figure*}
            \centering
            \includegraphics[width=\linewidth]{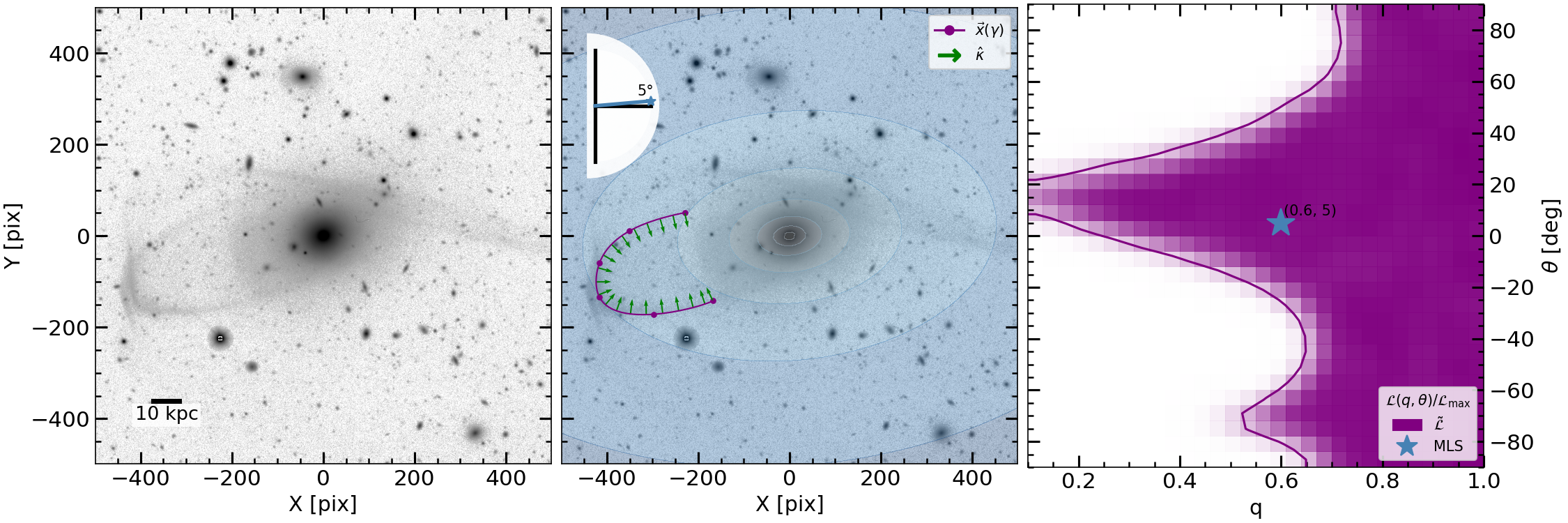}
            \caption{%
                IC169: Likelihood inference from a single stellar stream. %
                Colors, symbols, and panel arrangement follow \autoref{fig:ESO186-063}. %
                In this case, the selected MLS for the halo equipotentials is $(q, \theta)=(0.6,5\degree)$. %
            }
            \label{fig:IC169}
        \end{figure*}

        \autoref{fig:IC169} shows another example of a turning-point stream in a different host galaxy, the spheroid-dominated galaxy IC~169. %
        Although the host galaxy of the stream is different from UGC~8717, the stream morphology is analogous to that of UGC~8717: a compact arc confined to one side of the host, and two relatively straight, lower-curvature  segments pointing roughly toward the host galaxy. %
        The likelihood distribution in the right panel shows a similar structure to that of  UGC~8717 in \autoref{fig:UGC8717}, despite the fact that the IC~169 stream is located at a greater galactocentric radius than the UGC~8717 stream (see scale bars in the left panels). %
        As in the case for the UGC~8717 stream, the constraints from the IC~169 stream are weakest when the halo orientation aligns with the stream's symmetry axis (in this case $\theta\approx10\degree$, see purple region), while other orientations provide tighter limits. %
        We discuss this further in \autoref{subsec:sensitivity}. %

	\subsection{Greater Circles, Lesser Constraints}\label{sec:results:galaxy_weak}

        We now turn to systems in which the stream morphology provides little constraint on the projected halo geometry. %
        These streams exhibit a broad, ring-like morphology that in projection approximate great circle streams. Note that true great circles can also look like tight turns (see \autoref{sec:results:galaxy_Ushape}) in projection. %
        The ring-like streams presented here can appear as partial arcs, complete loops, or even multiply-wrapped loops. %
       
        \begin{figure*}
            \centering
            \includegraphics[width=\linewidth]{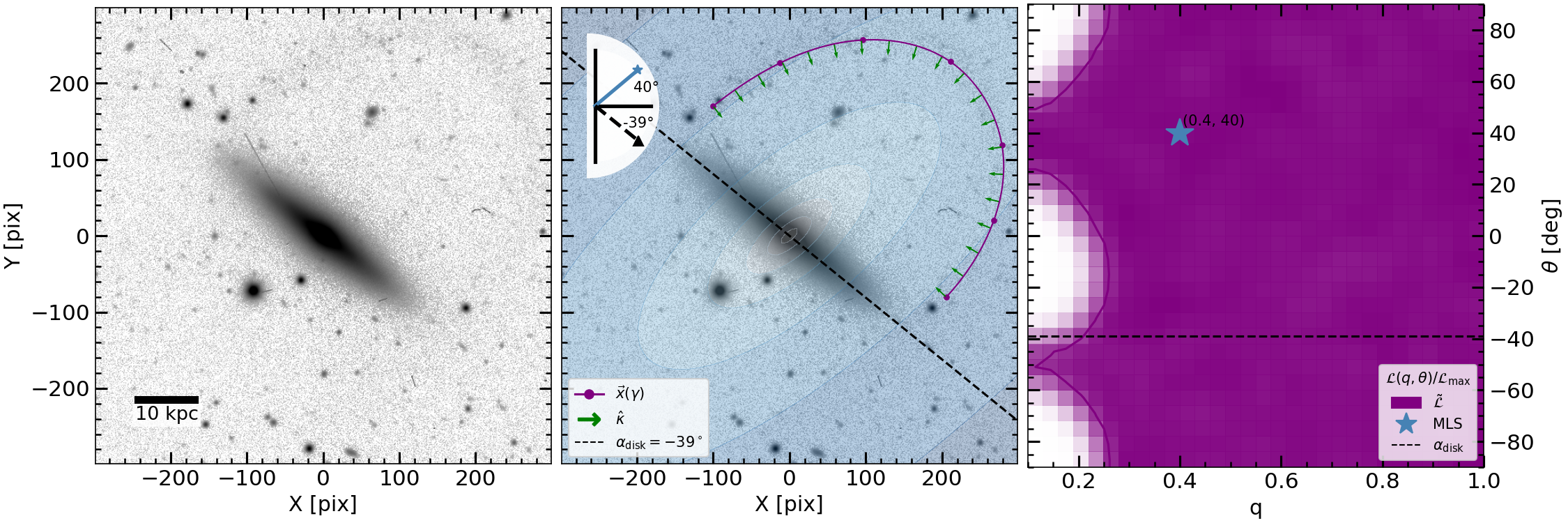}
            \caption{%
                ESO413-020: Likelihood inference from a single stellar stream. %
                Colors, symbols, and panel arrangement follow \autoref{fig:ESO186-063}. %
                In this case, the disk orientation is $\alpha_{\rm disk}=-39\degree$, and the selected MLS for the halo equipotentials is $(q,\theta)=(0.4,40\degree)$. %
            }
            \label{fig:ESO413-020}
        \end{figure*}

        \begin{figure*}
            \centering   
            \includegraphics[width=\linewidth]{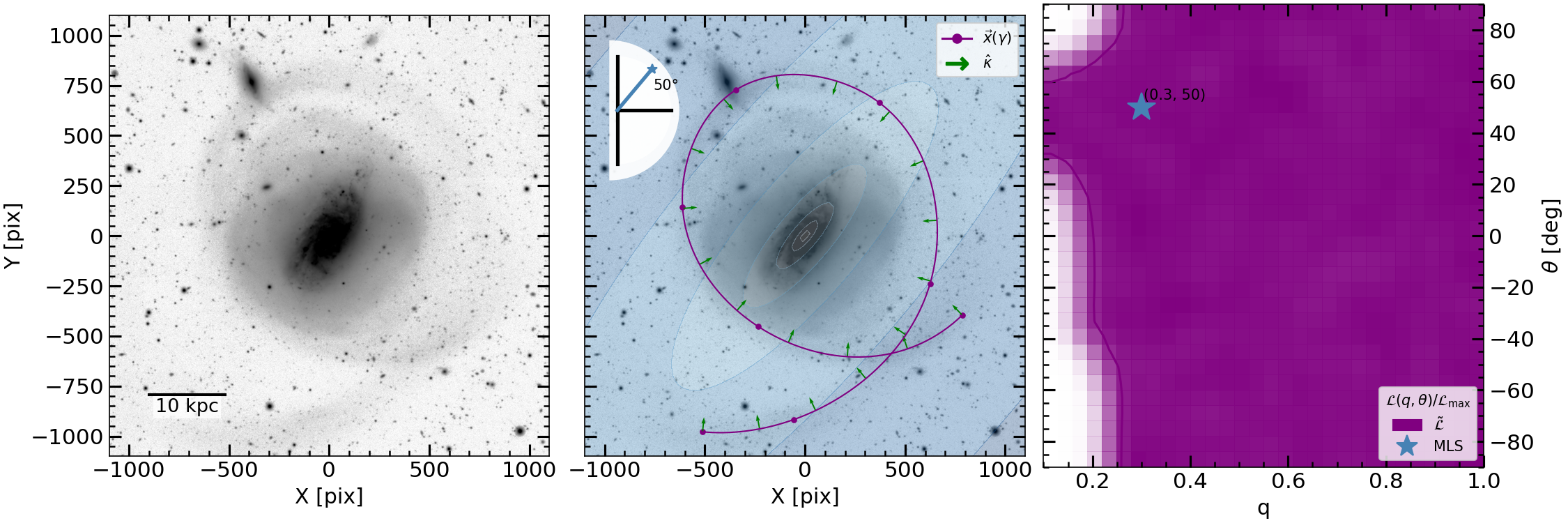}
            \caption{%
                NGC1084: Likelihood inference from a single stellar stream. %
                Colors, symbols, and panel arrangement follow \autoref{fig:ESO186-063}. %
                In this case, the selected MLS for the halo equipotentials is $(q,\theta)=(0.7,-50\degree)$. %
            }
            \label{fig:NGC1084}
        \end{figure*}

        \autoref{fig:ESO413-020} shows the curvature-based inference for the stellar stream around ESO~413--020. %
        The host is an edge-on disk galaxy with a well-defined major axis, marked by the dashed line. %
        The stream traces a wide, shallow arc resembling a half-ellipse on one side of the galaxy, and its curvature direction varies slowly along the track, with no sharp turning point. %
        We note that this stream can be considered an extreme case of a turning-point stream, as it has two lower-curvature segments on either side of the turning point, with the stream confined to one side of the galaxy.  %
      
        For the ESO~413--020 stream, most halo shape parameters that are not extremely flattened, yield an acceleration direction which aligns ($\varphi<90\degree$) with the curvature direction at all points along the stream. %
        As a result, the normalized likelihood surface in the $(q,\theta)$ plane is dominated by a broad plateau (right panel, \autoref{fig:ESO413-020}). %
        Only for strongly flattened halos ($q<0.25$) at certain orientations does a non-negligible fraction of reference points violate the $\varphi<90\degree$ criterion, producing the small excluded regions at low likelihood (see white regions). %
        
        \autoref{fig:NGC1084} shows an even less constraining case: the stellar stream around galaxy NGC~1084, which wraps around the host galaxy. %
        At nearly every reference point along the ridgeline (see green points in the middle panel), the curvature vector points approximately toward the galaxy center, and the projected acceleration vector does the same for a wide range of halo flattenings and orientations. %
        The condition $\hat{\boldsymbol{\kappa}}\cdot\hat{\mathbf{a}}_{xy}>0$ is therefore satisfied across most of parameter space. %
        As a consequence, the likelihood map is essentially a broad maximum (colored by deep purple) extending from near-spherical to very flattened halos with very little dependence on orientation. %
        Only extremely flattened models ($q<0.2$) generate enough local misalignment to be ruled out, so the marked \gls{mls} is best interpreted as one representative point on a large plateau rather than a unique best-fit solution. %

        The systems presented in this subsection illustrate a fundamental limitation of curvature-based inference for ring-like streams. %
        When the projected morphology closely follows a ring-like arc, the curvature direction is compatible with most acceleration direction over a wide range of halo model, and the method cannot strongly discriminate between halo geometries. %
        In contrast to turning-point streams, ring-like streams naturally satisfy the alignment condition across most of the $(q,\theta)$ plane,  leading to broad likelihood plateaus and weak constraints. %

        We present the remaining systems in \autoref{app:other_results} where the same general behavior recurs across similar morphologies. %

\section{Discussion}\label{sec:discussion}

    In this Section, we discuss the sensitivity of our constraints to stream geometry (\autoref{subsec:sensitivity}), the limitations of our method (\autoref{sec:discussion:limitations}), the relationship between inferred halo orientations and stellar disks in our sample (\autoref{subsec:disk}), and the future prospects for using the curvature method (\autoref{subsec:future}). %

    \subsection{Sensitivity Analysis}\label{subsec:sensitivity}

        Our likelihood is driven by a geometric test applied at each reference point: models are favored when the local curvature direction $\hat{\boldsymbol{\kappa}}$ 
        points within $90\degree$ of the predicted in-plane acceleration $\hat{\mathbf{a}}_{xy}$ (i.e., $\hat{\boldsymbol{\kappa}}\cdot\hat{\mathbf{a}}_{xy}>0$). %
        The constraining power of a stream is governed by the rate at which the in-plane acceleration vector and local curvature direction change along a stream track: violations of $\varphi<90\degree$ in the alignment test occur more easily where one direction has rotated appreciably while the other has not. %
        The more out of sync the two vectors are, the stronger the constraints become. %

        \begin{figure*}
            \centering
            \includegraphics[width=\linewidth]{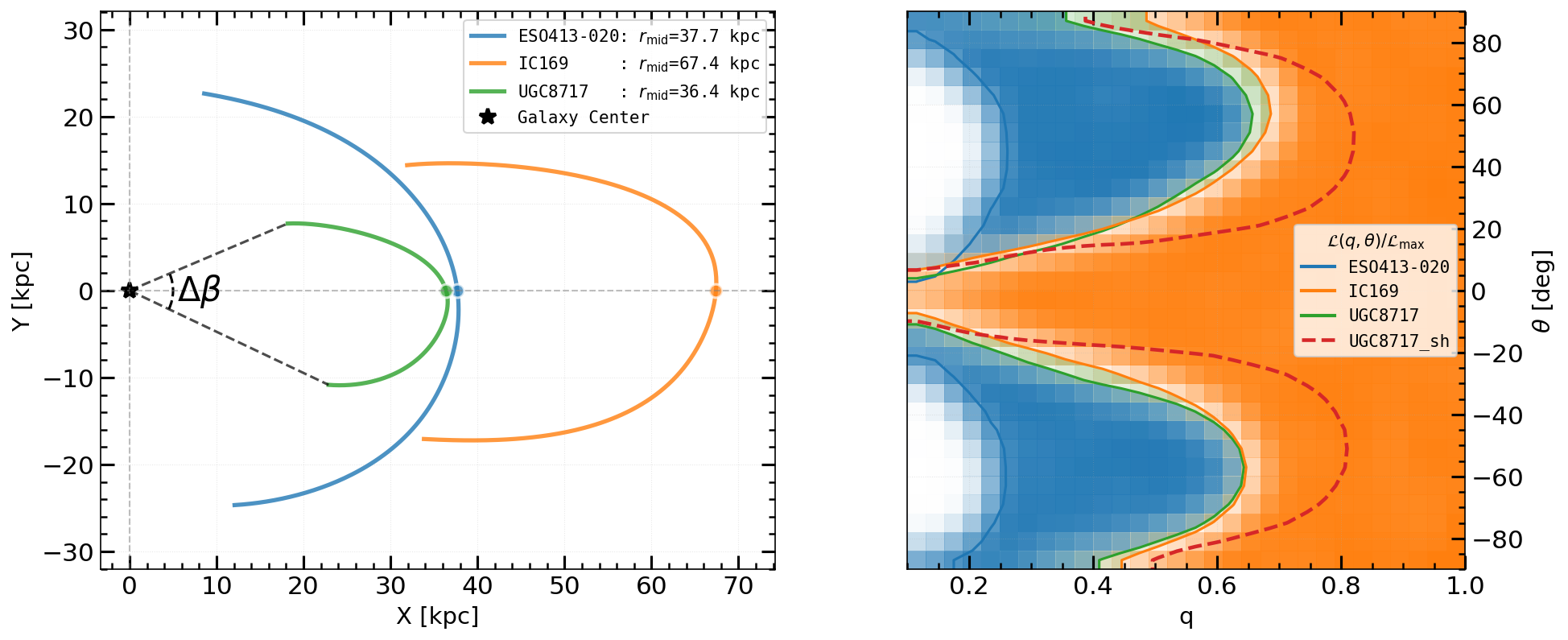}
            \caption{%
                Comparative analysis of three systems exhibiting turning-point stream morphologies: ESO413-020 (blue), IC169 (orange), UGC8717 (green). Left: Projected stream ridgeline in physical coordinates (kpc). The systems are rotated to a common frame where the geometric midpoint of each stream (indicted by a dot) aligns with the positive x-axis, placing the host galaxy center at the origin. The black dashed line shows the azimuthal span $\Delta\beta$, for the stream in UGC 8717. The galactocentric distances of the midpoint, $r_{\rm mid}$, are listed in the legend. Physical scales are derived from spectroscopic redshift from SIMBAD. 
                Right: Normalized likelihood distributions, %
                $\mathcal{L}/\mathcal{L}_{\max}$. %
                We also show the result (green dashed), obtained by shifting the stream ridgeline of UGC8717 to the physical location of IC169. %
                This visualization highlights the roles of the importance of azimuthal sweep, the turning-point opening angle and the galactocentric distance. %
                Tighter turns generically admit fewer halo geometries, providing tighter constraints. %
            }
            \label{fig:3Ushape-stream}
        \end{figure*}

        To investigate the geometric factors that control this mismatch between the local curvature and in-plane acceleration, we place three turning-point streams into a common reference frame in \autoref{fig:3Ushape-stream} (left), scaling them by their physical galactocentric distance and rotating them so that the midpoint of each ridgeline lies on the positive $x$-axis. %
        Because these morphologies are approximately symmetric about the $x$-axis, the resulting constraints (in the right panel) show a characteristic ``W''-shaped pattern, rotated by $90\degree$ in the $(q,\theta)$ plane. %
        The least constrained regions occur when 
        $\theta$ is either aligned ($\theta\approx0\degree$) or perpendicular ($\theta\approx\pm90\degree$) to the midpoint direction. 
        This is because the planar acceleration fields at both orientations are symmetric about the $x$-axis, matching the approximate symmetry of the stream, which reduces the geometric tension in the alignment test. %
        When $\theta\approx0\degree$ for an extremely flattened halo (e.g., $q=0.15$), the planar acceleration field is dominated by its $y$-component and directed toward the $x$-axis. %
        Because $\hat{\boldsymbol{\kappa}}$ points approximately toward the $x$-axis, the angle $\varphi$ between $\hat{\boldsymbol{\kappa}}$ and $\hat{\mathbf{a}}_{xy}$ remains acute at all reference points, so $\hat{\boldsymbol{\kappa}}\cdot\hat{\mathbf{a}}_{xy}>0$ is satisfied everywhere. %
        Consequently, the halo flattening is unconstrained at this orientation. %
        When $\theta\approx\pm90\degree$, the major axis is perpendicular to the midpoint direction, producing an acceleration field with a strong $x$-component. %
        For the most extreme flattenings, the acceleration direction at the lower-curvature parts at the end of the streams 
        deviates sufficiently from the local curvature direction that $\varphi$ exceeds $90\degree$, excluding those configurations while still permitting moderate values of $q$. %
        At intermediate orientations, the acceleration field breaks this mirror symmetry. %
        The curvature and acceleration directions then evolve out of sync along the track, producing tighter constraints. %

        A key analytic insight underlies the three stream comparisons that follow in \autoref{fig:3Ushape-stream}. %
        The acceleration direction depends only on the local position azimuth $\beta$ of the reference point along the stream, not on its radial distance. %
        For $\mathbf{r} = (r\cos\beta,\, r\sin\beta)$, the acceleration orientation is
        \begin{equation}\label{eq:acc_angle}
            \tan \alpha_a(\beta, q, \theta) = \frac{c_{xy} + c_{yy}\tan\beta}{c_{xx} + c_{xy}\tan\beta} \, ,
        \end{equation}
        where the matrix elements ($c_{xx}, c_{yy}, c_{xy}$) from \autoref{eq:methods:potential:acc_general} are fixed for a given $(q,\theta)$, and this dependence is monotonic: $\alpha_a$ advances in the same sense as $\beta$. %
        The rate at which $\beta$ changes along the ridgeline,
        \begin{equation}
            \frac{d\beta}{ds}=\frac{\mathbf{r}\times\hat{\mathbf{t}}}{r^2} \, ,
        \end{equation}
        depends on the stream morphology and galactocentric distance. %
        The span $\Delta\beta$ swept by the turning-point stream therefore sets the angular range of acceleration directions sampled, while the curvature-direction span is controlled by the intrinsic morphology and viewing angle. %

        First, comparing UGC~8717 with the same ridgeline translated to the galactocentric distance of IC~169 (green dashed contour in \autoref{fig:3Ushape-stream}) isolates the effect of distance. %
        The translation compresses $\Delta\beta$ without altering the curvature-direction span. %
        For a turning-point stream, whose curvature direction already sweeps a large angular range, the acceleration direction now spans only a narrow range by comparison. %
        The resulting constraint contour shifts toward higher $q$, indicating that greater galactocentric distance strengthens the constraints for turning-point streams by amplifying the mismatch between the curvature and acceleration directions. %

        Second, comparing between UGC~8717 and IC~169 (orange contour)  tests the role of the position-azimuth span $\Delta\beta$. %
        Despite different galactocentric distances ($r_{\rm mid}$ in the legend), both streams sweep comparable $\Delta\beta$ and share similar opening morphologies. %
        Because the acceleration direction depends primarily on $\beta$ (\autoref{eq:acc_angle}), the two streams probe nearly the same set of acceleration directions and curvature directions, producing almost identical constraint contours. %

        Third, comparing UGC~8717 with ESO~413-020 (blue contour) demonstrates the impact of the turning-point opening angle. %
        ESO~413-020 has a wider, more open turn, so the curvature and acceleration directions rotate more nearly in step along the track; the geometric test is therefore easier to satisfy, and the constraints are weaker. %
        UGC~8717, by contrast, has a narrower turn, so the two direction fields evolve less synchronously, leading to stronger constraints. %

        We further note that the observed opening angle of a turning-point stream is not purely an intrinsic properties of stream's progenitor, but is itself shaped by the viewing geometry. %
        A three-dimensional turning arc viewed at high inclination (edge-on) is projected into a narrower opening than the same arc viewed face-on. %
        This projection effect simultaneously compresses the opening angle and increases the rate at which the curvature direction $\hat{\boldsymbol{\kappa}}$ rotates through the turn, amplifying the mismatch with the acceleration-direction field on both counts. %
        The viewing angle and opening angle are therefore coupled, and this coupling also accounts for the contrast between the nearly edge-on UGC~6397 and the more face-on ESO~413-020 noted in \autoref{sec:results:galaxy_good} and \autoref{sec:results:galaxy_weak}. %
        
        In summary, the constraining power of a stream is not determined by any single geometric property, but by the degree of change between the curvature-direction field and the acceleration-direction field along the track. %
        The viewing angle controls the projected curvature field; the galactocentric distance and stream morphology together control the acceleration-direction field; and the opening angle of a turning point determines how much intrinsic curvature-direction span the stream carries. %
        Tighter constraints arise whenever these factors conspire to make the two direction fields evolve out of sync, generating $\varphi>90\degree$, which rules out incorrect halo models. %

    \subsection{Considerations of the Method}\label{sec:discussion:limitations}

        Here, we discuss various considerations for the methodologies described in this work. %

          \subsubsection{Other debris features}
            Our inference framework relies on the coherence condition introduced by \citetalias{Nibauer+:2023:Constraining} (see also \autoref{sec:methods:likelihood:combining}). %
            This assumption requires that the local acceleration possesses a component pointing toward the local center of curvature. %
             %
            While our curvature-based method can be applied to any annotated ridgeline for which the curvature is well defined, the resulting constraints are physically meaningful only when the underlying feature satisfies the coherence condition. 
            Because astronomical observations only provide a single snapshot in time ambiguities can arise. %

            \begin{figure*}
                \centering
                \includegraphics[width=\linewidth]{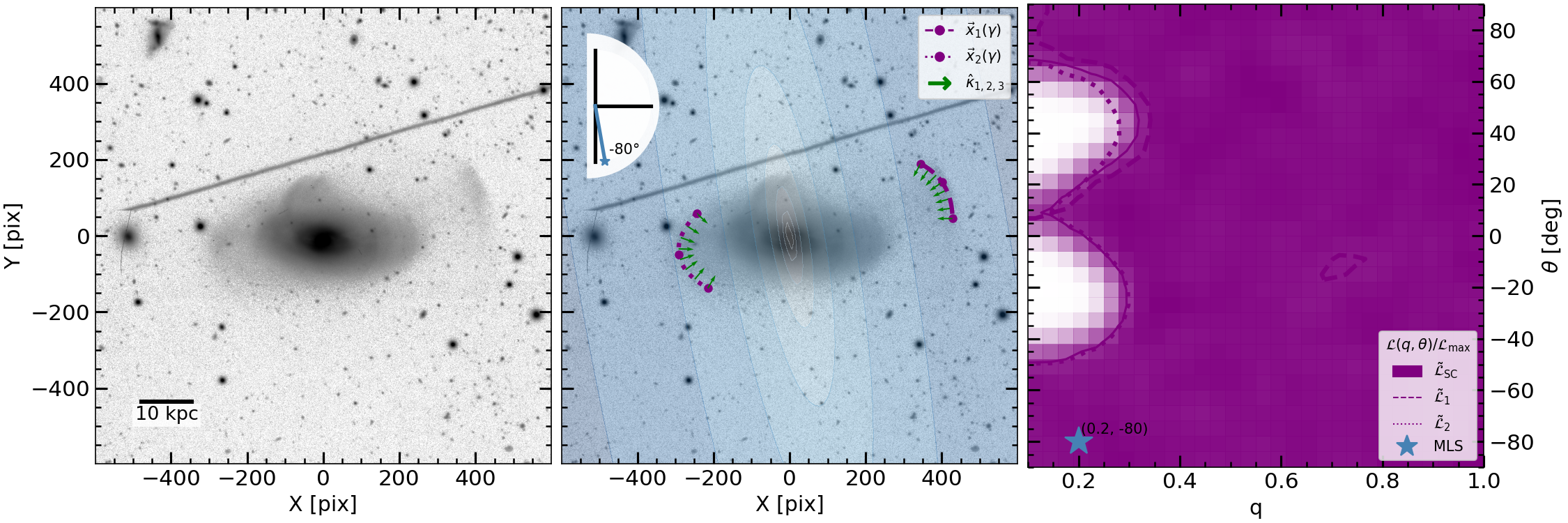}
                \caption{%
                    NGC7506: Likelihood inference from multiple shell-like tidal features. %
                    Note: the straight streak through the image is an observational artifact. %
                    Left: Raw image with a 10 kpc scale bar. %
                    Middle: The fitted stellar tidal features ($\vec{x}_1(\gamma)$, purple
                    dashed; $\vec{x}_2(\gamma)$, purple dotted), overlaid on the halo equipotentials (light blue contours) of the selected MLS marked by the star in the right panel. %
                    The unit-circle inset shows the MLS halo major-axis orientation $\theta=-80\degree$. %
                    Right: Normalized profile likelihood $\mathcal{L}(q,\theta)/\mathcal{L}_{\rm max}$ over halo projected flattening $q$ and orientation $\theta$. %
                    Purple shading indicates the feasible region (darker = higher likelihood) for the combined inference $\tilde{\mathcal{L}}_{\rm SC}$; the darkest plateau represents equally likely maximum-likelihood models, among which the star marks one representative MLS at $(q, \theta)=(0.2,-80\degree)$. %
                    Purple dashed/dotted lines: per-feature feasible boundaries $\tilde{\mathcal{L}}_{1,2}$. %
                    For both, the feasible region lies to their right (higher $q$), while the region to the left is excluded. %
                }
                \label{fig:NGC7506}
            \end{figure*}

            The SSLS catalog contains tidal features beyond stellar streams, including shell-like structures and candidates that may instead be tidal tails \citep{Miro-Carretero+:2024:SSLS}. %
            \autoref{fig:NGC7506} shows NGC~7506, which contains two prominent shell-like tidal features rather than stellar streams. %
            Shells are typically produced by low-angular-momentum accretion events, where debris moves on nearly radial trajectories and phase-wraps to form caustic surfaces at orbital apocenters \citep{Sanderson+Helmi:2013:AnalyticalPhasespaceModel,pop2018}. %
            The caustic surfaces are formed by the superposition of many stars at their orbital apocenters and do not obey the coherence condition. %

            Geometrically, the observed shell segments in NGC~7506 appear as wide-opening, apocentric-turn-like arcs.
            To demonstrate an example of the curvature analysis applied to shells, in \autoref{fig:NGC7506} (right), we present the likelihood analysis, which indicates that these morphologies provide very little constraints on the halo shape. %
            The right-hand shell ($\hat{x}_1$) excludes  $\theta\approx0\degree$--$60\degree$ for strongly flattened halos ($q \approx 0.1$--$0.35$), while the left-hand shell ($\hat{x}_2$) additionally rules out $\theta\approx -50\degree$--$0\degree$ in the same $q$ range. %
            The combination of the two segment constraints disfavors very flattened halos ($q \lesssim 0.35$), but cannot further constrain orientation or distinguish moderately flattened from spherical halos. %
            The likelihood should be interpreted only as an exploratory diagnostic rather than a robust halo constraint derived from a true stream ridgeline. %

            The weak constraining power can be understood since the wide-opening shell morphology is analogous to the circular-like streams discussed in \autoref{sec:results:galaxy_weak}: shells generally trace the equipotential surfaces of the host halo, so the local curvature vector $\hat{\boldsymbol{\kappa}}$ points approximately inward, naturally aligning with the gravitational acceleration $\hat{\mathbf{a}}_{xy}$, such that the alignment condition $\hat{\boldsymbol{\kappa}}\cdot\hat{\mathbf{a}}_{xy}>0$ is satisfied for most halo geometries.  %
 
            While our purely geometric approach for analysing shells is not constraining, kinematic data offer a pathway to stronger constraints. %
            \citet{Sanderson+Helmi:2013:AnalyticalPhasespaceModel} showed that tidal shells are caustics in phase space whose curvature in the radius-velocity plane is inversely proportional to the local radial acceleration. 
            If we can obtain line-of-sight velocity measurements for shells we could independently constrain the gravitational potential, even for flattened halos  \citep[see e.g.,][]{escala2022}. %
            
            Finally, we note that the SSLS catalog also contains features that may be tidal tails rather than stellar streams. %
           Tidal tails are produced in major mergers or strong interactions, where resonances between the orbital and internal rotational angular velocities drive the formation of extended tidal structures 
            \citep{Toomre+Toomre:1972:GalacticBridgesTails, Dubinski+:1996:UsingTidalTails, Barnes+Hibbard:2009:Identikit1Modeling}. %
            Unlike streams that trace a single coherent orbit, galactic tidal tails consist of a broad distribution of debris with divergent dynamical fates: the inner regions often lose energy and fall back into the host galaxy (forming shells), while the ends of the tails may become unbound and escape. %
            This complex, non-equilibrium configuration violates our foundational assumption that a feature represents a coherent orbit (\autoref{sec:discussion:limitations}) and curvature-based inference is therefore ill-suited for tidal tails. %
            Some ambiguity exists in classifying a feature as a stellar stream versus a tidal tail, since not all stellar streams orbit far from or disconnected from the baryonic disks of their host in projection  \citep[see also discussion in][]{Sakowska+:2025:StellarStreamsDwarf}. 
            In practice, tidal tails are usually morphologically distinct, appearing broader, hotter, or connecting massive companions, and can be excluded during the sample selection. %
            Photometric color information may also aid this classification, as tidal tails are expected to share the colors of the host galaxy, whereas stellar streams typically originate from older satellite populations. %

        \subsubsection{Inferring the intrinsic halo shape}\label{sec:inferring_the_intrinsic_halo_shape}

            In this work, we constrain the projected flattening $q$ and the projected orientation $\theta$ of the halo potential. %
            To interpret these two-dimensional observables in the context of the intrinsic three-dimensional halo shape, we rely on the geometric derivation presented in \autoref{app:projection_proof}. %
            There, we prove that the projected axis ratio $q$ serves as a strict upper limit to the intrinsic short-to-long axis ratio ($c/a$) of a triaxial ellipsoid. %
            In other words, for any viewing angle, the observed flattening cannot be narrower than the intrinsic flattening ($q \ge c/a$). %
            To aid in interpreting these geometric effects, we provide an interactive visualization tool available in our code repository.\footnote{The visualization tool is available at: \url{https://github.com/wsr1998/ellipsoid-projection-bound}}. %

            This geometric inequality has important implications for our ability to constrain the halo's sphericity. %
            A measurement where the maximum allowed flattening is less than unity ($q_{\max} < 1$) would strictly imply $c/a < 1$, thereby ruling out a perfectly spherical halo. %
            In our current observational sample of 15 streams, the posterior distributions generally extend to unity ($q_{\min} < q \le 1$), meaning the data cannot definitively exclude spherical configurations. %
            However, this absence of strict exclusion reflects the limited geometric diversity of our modest sample rather than a fundamental methodological flaw. %
            The specific orbital configurations required to bound $q_{\max}$ strictly away from unity , for example, streams with highly asymmetric trajectories are geometrically rare and are not represented in our current sample. %
            To verify that the method itself is capable of excluding spherical models, we ran tests using mock streams generated with \texttt{galax} in known flattened potentials and successfully recovered posteriors bounded strictly away from unity ($q_{\max} < 1$), demonstrating that the method can rule out spherical halos given sufficiently constraining orbital geometries. This was also shown directly in \citetalias{Nibauer+:2023:Constraining}. %
            In \citetalias{Nibauer+:2023:Constraining} it is shown that the 3D halo configuration can be constrained by assuming a strictly continuous line-of-sight path along a stream. However, this yields constraints only for highly informative stellar streams like those which change concavity. We do not observe these in our sample, so our current constraints would only be made weaker by allowing for a more flexible halo model. %
            
            Fully breaking the degeneracy between the projected ($q, \theta$) and intrinsic 3D shape requires additional information or modeling assumptions. %
            We identify three primary pathways to resolve this degeneracy in future work without sacrificing robustness in modeling. %
            First, detecting streams with distinct, highly asymmetric orbital configurations \citep[such as ``the dog leg stream''][]{Amorisco+:2015:DwarfGalaxysTransformation} could provide tighter geometric limits that rule out spherical models. %
            Second, while the streams in our sample are mostly too low surface brightness to obtain kinematics along the streams from  globular clusters, planetary nebulae, or co-addition of surface brightness fluctuations, incorporating kinematic stream information would help probe the three-dimensional phase-space structure \citep[see e.g.,][]{Fardal+:2013:InferringAndromedaGalaxys,Foster2014,Toloba+:2016:NewSpectroscopicTechnique, Pearson+:2022:MappingDarkMatter,Hughes+:2023:NewVelocityMeasurements, Muller+:2025:MUSEObservationsDwarf, Valenzuela2026}. %
            Third, comparing our observations with tailored galaxy reconstructions or high-resolution hydrodynamic simulations could offer stronger priors on the intrinsic shape distribution \citep[e.g.,][]{agora2014}. %

        \subsubsection{Track Continuity and Observational Completeness} \label{sec:discussion:continuity_and_length}

            In \autoref{sec:results:galaxy_good}, we presented UGC~6397 (\autoref{fig:UGC6397}) as a strongly constraining system whose stream is partially obscured by the host galaxy. %
            For such visually interrupted streams, we compare two reconstruction strategies: one assumes that the stream is physically continuous and interpolates across the obscured region, while the other uses only the visible segments. %
            As shown by UGC~6397 and the additional cases in \autoref{appendix:visually_interrupted_streams}, the two strategies generally provide similar constraints, but can diverge at specific orientations. %
            This discrepancy has two origins. %
            First, the interpolated segment itself may carry significant geometric tension -- in UGC~6397, for instance, this interpolated segment is nearly straight and passes close to the galactic center, where the acceleration direction span is much larger than the curvature direction span, and therefore introduces more constraints as mentioned in \autoref{subsec:sensitivity}. %
            Second, when the visible portions are fitted as independent segments, the spline reconstruction near the segment ends can produce slightly different curvature directions from those obtained in the continuous reconstruction. %
            Because \Potamides~ is intended to operate as a rapid geometric filter, a conservative choice is to adopt the union of the allowed halo shapes by the two reconstruction strategies, so that no viable configuration is excluded purely because of the track-reconstruction uncertainty. %

            When a stream is plausibly continuous, in this paper we construct both a continuous annotation and a segmented one, while requiring the curvature direction near the segment ends to remain as consistent as possible between the two strategies. %
            When interpolating across the obscured region, we also avoid introducing changes in concavity, i.e., inflection points, because such features would flip the curvature direction by $180\degree$ and could spuriously bias the result. %
            This is a conservative treatment, since some Milky Way streams exhibit genuine discontinuities due to resonances, perturbations from satellites or substructure \citep{sesar2016,Li2021,bonaca2019, dillamore2022}. Note however, we do not include time-dependent perturbers to streams in this paper. %

            A related issue is observational incompleteness caused by surface-brightness limits, which often make the faint wraps difficult to detect \citep[although see recent example in M83:][]{Bell+:2026:LowmassStructuredStellar}. %
            Improved imaging, e.g. from upcoming and ongoing space missions, or more advanced host-galaxy subtraction may reveal additional stream segments that strengthen the constraints -- not by adding length (see \autoref{fig:IC160}), but by revealing additional stream segments that sweep a wide range of curvature directions. %
            In this sense, the current result is conservative: missing faint structure is more likely to add constraining power than to overturn the constraints already inferred from the visible ridgeline. %

    \subsection{A Comment on Disk Alignment}\label{subsec:disk}

        For three streams in our sample (ESO~186-063: \autoref{fig:ESO186-063},  UGC~8717: \autoref{fig:UGC8717}, and ESO413-020: \autoref{fig:ESO413-020}), the likelihood surfaces are consistent with any degree of halo flattening when the projected major axis is either aligned with or perpendicular to the disk. %
        
        This can be interpreted as evidence that the baryonic disk shapes the inner gravitational field, in line with hydrodynamical $\Lambda$CDM simulations where baryonic feedback drives the inner halo into alignment with the disk \citep{Bailin+:2005:InternalAlignmentHalos, Prada+:2019:Auriga}. 
        
        Although this is an exciting prospect, this result could also simply be due to our low sample size and viewing these halos at a random orientation. %
       In \autoref{subsec:sensitivity}, we showed that for a turning-point stream the halo flattening $q$ is unconstrained when the projected halo major axis aligns nearly with the stream's morphological symmetry axis (i.e. where the stream turns around). %
        The  fact that all halo flattenings are allowed in this region,  
        is therefore consistent with the stream's symmetry axis happening to fall close to the disk major axis (or  perpendicular to the disk major axis), placing $\alpha_{\rm disk}$ in the geometrically unconstrained region of the $(q,\theta)$ plane. %
        
        A random-orientation argument supports this interpretation. %
        If the stream symmetry axis were uniformly distributed, the probability of falling within $\pm 10\degree$ of either $\alpha_{\rm disk}$ or  perpendicular to it would be $4\times10\degree / 360\degree \approx 22\%$. %
        Three of the fifteen systems in our sample ($20\%$) satisfy this condition, consistent with this expectation. %
        
        A selection effect likely compounds this geometric coincidence, since a stream for which the turning point projects near the disk normal sits in the low-surface-brightness sky above or below the disk and is therefore easier to detect than one projected onto the bright disk light. %
        Because our targets were drawn from visual identification of tidal features, this bias favors configurations in which the morphological symmetry axis lies close to the disk normal. %

        We emphasize that these geometric and selection-based considerations do not exclude a  physical alignment between the inner halo and the baryonic disk. %
        \Potamides~ currently employs a single-component potential whose inferred $(q,\theta)$ describe the total effective gravitational field, so it cannot separately attribute an alignment signal to the disk or the halo. %
        Disentangling a true alignment signature from geometric and observational bias will require multi-component modeling and a larger, less visually biased sample. %

    \subsection{Future Prospects}\label{subsec:future}
        
        In this work, we have modeled the total gravitational field using a single flattened logarithmic potential,  which implicitly encompasses the contributions from both the dark matter halo and each galaxy's baryonic components. %
        In future work, we will refine our gravitational potential model by explicitly incorporating a baryonic disk component (e.g., a Miyamoto-Nagai disk), following the framework introduced in \citetalias{Nibauer+:2023:Constraining}. %
        This multi-component approach will be critical in order to test whether disks induce flattenings of inner halos, which in turn allows us to test predictions from $\Lambda$CDM and alternative dark matter models \citep[e.g.,][]{Giocoli+:2026:AIDATNGProject3D}. %
        If we explicitly include baryonic components in the gravitational model, that will further enable us to capture non-spherically symmetric effects,  such as the influence of massive satellites \citep{Weerasooriya+:2025:DancingStreamsMerging, Brooks+:2025:LMCCallsMilky}, and to test whether stellar streams preferentially align with the stellar disk. %
       
        On the observational side, our results carry a clear implication for future target selection. %
        The most constraining systems are not necessarily the longest or oldest streams, but those whose projected morphology drives the strongest mismatch between curvature and acceleration directions -- for example, streams viewed nearly edge-on or those with sharp turning points. %

        The quality of the input data can further boost the geometric constraint: advanced image processing techniques, such as the model-based host-galaxy subtraction demonstrated for UGC~8717 by \citet{Sola+:2025:STRRINGS}, can extend the visible arc of a stream into regions that sweep a wide range of curvature directions, thereby strengthening the constraints. %

        Even in its current form, \Potamides~ serves as a rapid geometric filter. It is not only several orders of magnitude faster than forward modeling techniques, but can simultaneously exclude broad regions of the $(q, \theta)$ space, and provide informed priors that reduce the computational expense of subsequent detailed forward modeling \citep[e.g.,][]{Fardal+:2013:InferringAndromedaGalaxys,Pearson+:2022:MappingDarkMatter,Nibauer+Pearson:2025:TestingDarkMatter,Chemaly+:2026:HierarchicalBayesianInference}. %

       The sample presented in this paper is based on ground-based data \citep{Martinez-Delgado+:2023:SSLS,Miro-Carretero+:2024:SSLS} and, thus, is limited by the surface brightness cutoffs of the observations and the ``mosaicking''  to compose the image cutouts, which removes some of the trace of faint features associated with the streams.        
       The full potential of  \Potamides~ will be realized with the large stream samples expected from \textit{Euclid} \citep{Racca+:2016:EuclidMissionDesign}, the Vera C.\ Rubin Observatory \citep{LSST+:2019:LSSTScienceDrivers}, the Nancy Grace Roman Space Telescope \citep{Spergel+:2015:WideFieldInfrarRedSurvey}, and ARRAKIHS \citep{Guzman+:2022:DUNESARRAKHISSpace}. %
       With such samples, we can move the method beyond object-by-object case studies and toward population-level constraints, which can help average over projection effects and orbital diversity and thereby provide sharper tests of baryonic effects, $\Lambda$CDM, and alternative dark matter models \citep{Giocoli+:2026:AIDATNGProject3D}. %
      These datasets will also enable the detection of fainter debris structures farther from the baryonic disks. %

\section{Conclusion} \label{sec:conclusion}

    In this work, we presented \Potamides, a curvature-based inference framework developed from \citetalias{Nibauer+:2023:Constraining}, which uses only the projected ridgeline morphology of extragalactic stellar streams to place probabilistic constraints on the projected flattening and orientation of the host potential. %
    
    We applied this method to a population of tidal features for the first time, using 15 stellar streams from the Stellar Stream Legacy Survey \citep{Martinez-Delgado+:2023:SSLS, Miro-Carretero+:2024:SSLS}. We have showed that two-dimensional stream morphology alone can exclude substantial regions of the projected halo geometry. 
    Below we list our main conclusions from this work:

\begin{enumerate}[leftmargin=1em]
    \item {\bf Observational constraints on halo geometry.} %
    The constraining power varies significantly across the sample. %
    Systems hosting streams with tight opening or high-inclination wrapping loops exclude large regions of the halo flattening and halo orientation $(q, \theta)$ plane. %
    Some turning-point streams 
    restrict $q$ to $\gtrsim 0.7$ at orientations away from their respective symmetry axes. %
    In contrast, ring-like great circle streams satisfy the geometric alignment test for nearly all halo models and therefore place negligible constraints. %
    Overall, the well-constrained systems in our sample favor mildly flattened to spherical halo shapes in projection. %
    
    \item {\bf Origin of the constraining power.} %
    The degree of mismatch 
    between the curvature-direction field and the acceleration-direction field along the stream ridgeline governs 
    the constraining power. 
    Three geometric factors control the degree of mismatch between these two vectors: %
    (i) the viewing inclination, which sets how rapidly the curvature direction varies along the observed stream ridgeline; %
    (ii) the galactocentric distance govern the rate of change of acceleration direction; %
    (iii) the opening angle of any turning point, which determines the intrinsic angular span of curvature directions the stream carries. %
    These factors are not independent: the viewing angle modulates the opening angle through projection. %
    Tighter constraints arise whenever these factors make the two direction fields evolve out of sync, generating sign flips on the geometric alignment test that eliminate incompatible halo models. %
    
    \item {\bf Geometric bound on intrinsic flattening.} %
    Collisionless $\Lambda$CDM only simulations predict triaxial halos \citep[][]{Jing+Suto:2002:TriaxialModelingHalo}, however, for all halos analysed in this paper, the posterior distributions extend to $q = 1$, and we cannot rule out spherical configurations. %
    Given the expectation of triaxiality and the ability of this method to exclude sphericity for certain stream classes, it is notable that none of the prominent SSLS streams excludes spherical halo configurations. %
    The projected axis ratio $q$ serves as a strict upper bound on the intrinsic short-to-long axis ratio of a triaxial halo, $q \ge c/a$. %
    A measurement with $q_{\max} < 1$ would therefore strictly exclude a spherical halo. %
    SIDM predicts more spherical halos \citep[e.g.,][]{Dave+:2001:HaloPropertiesCosmological,Despali+:2022:ConstrainingSIDMHalo,Giocoli+:2026:AIDATNGProject3D}, and future larger samples of extragalactic stellar streams, will allow us to test the SIDM predictions of spherical halos and the $\Lambda$CDM prediction of non-spherical halos in more detail. %

    \item {\bf Disk-halo alignment.} Three of the stellar streams presented in this paper show evidence that the baryonic disks shape the inner gravitational field. In the small sample size of 15 systems presented in this work, we show that this can be explained from a random orientation argument. With the upcoming large sample sizes of extragalactic streams from Euclid, ARRAKIHS, LSST, and Roman, we will be able to test if there is indeed a preference for disk-halo alignment in the inner parts of halos, as predicted from e.g., \citet{Chua+:2019:ShapeDarkMatter, Prada+:2019:Auriga,Giocoli+:2026:AIDATNGProject3D}. 

\end{enumerate}

The results presented in this work show that projected stream morphology can provide efficient and physically meaningful constraints on halo geometry, even in the absence of spectroscopic kinematic data. \Potamides~ is several orders of magnitude faster than forward modeling techniques typically used to model extragalactic stellar streams, and with 
future data we can extend our method to thousands of tidal features \citep[e.g.,][]{Racca+:2016:EuclidMissionDesign,LSST+:2019:LSSTScienceDrivers,Guzman+:2022:DUNESARRAKHISSpace}. This provides a complementary probe of DM halo shapes compared to gravitational lensing. By constraining the distribution of halo shapes across host galaxy properties and redshift, it will be possible to directly test fundamental and underexploited predictions of $\Lambda$CDM regarding dark matter halo shapes. %

\begin{acknowledgments}
    This work was supported by a research grant (VIL53081) from VILLUM FONDEN. %
    This work was also co-funded by the European Union (ERC, BeyondSTREAMS, 101115754) grant. %
    Views and opinions expressed are however those of the author(s) only and do not necessarily reflect those of the European Union or the European Research Council. %
    Neither the European Union nor the granting authority can be held responsible for them. %
    In preparing this manuscript, the authors used the latest and most capable paid subscription versions of three AI tools: Claude (\citeauthor{Anthropic2026}), ChatGPT (\citeauthor{OpenAI2026}), and Gemini (\citeauthor{Google2026}). %
    Specifically, we used Claude 4.5/4.6 Sonnet \& Opus, ChatGPT 5.1/5.2/5.3/5.4/5.5 Thinking \& Codex, and Gemini 3.0/3.1 Pro. %
    These tools were used to improve language clarity, search for relevant literature, troubleshoot code, enhance data visualizations, and draft documentation for our code repository. %
    The authors independently read and verified all cited sources, thoroughly reviewed all AI-assisted content, and take full responsibility for the manuscript. %
    S.W. would like to thank Jiang Chang, Keyu Lu and Chengye Cao for for their invaluable feedback and extensive discussions. %
    N.S. gratefully acknowledges support for this work from The Brinson Foundation through a Brinson Prize Fellowship. %
    J.N. is supported by a National Science Foundation Graduate Research Fellowship, Grant No. DGE 2039656. %
    Any opinions, findings, and conclusions or recommendations expressed in this material are those of the author(s) and do not necessarily reflect the views of the National Science Foundation. %
\end{acknowledgments}

\software{%
    \Potamides \citep{Potamides:2026},
    \package{astropy}[https://www.astropy.org] \citep{Astropy+:2013,Astropy+:2018,Astropy+:2022},  
    \package{coordinax}[https://coordinax.readthedocs.io/en/latest/] \citep{Starkman+:2026:Coordinax},
    \package{galax}[https://github.com/GalacticDynamics/galax] \citep{Starkman+:2024:GalacticDynamicsGalaxV002}, 
    \package{JAX}[https://docs.jax.dev/en/latest/] \citep{Bradbury+:2021:JAX}, 
    \package{matplotlib}[https://matplotlib.org]\citep{Hunter:2007:Matplotlib2DGraphics}, 
    \package{unxt}[unxt.readthedocs.io/] \citep{Starkman+:2025:UnxtPythonPackage}, 
    Claude (\citeauthor{Anthropic2026}), ChatGPT (\citeauthor{OpenAI2026}), Gemini (\citeauthor{Google2026}). 
}

\appendix
\section{Stream-track optimization details} \label{sec:appendix:optimizing_the_track}

    Human annotations provide an excellent starting point for characterizing stellar streams. %
    Visual inspection is effective at detecting faint, coherent stream structures and at extrapolating stream continuity. %
    However, hand--drawn annotations do not provide the smooth ridge line needed for dynamical inference. %
    Small-scale changes in concavity introduced by characterization points placed along the stream can significantly bias the curvature estimates that enter our likelihood analysis (see \autoref{eq:methods:likelihood}). %
    We address this by constructing an optimized representation of the stream track, a smooth spline that passes through the annotated points. %
    Because our inference depends directly on the \emph{direction} of the curvature vector along the track (\autoref{eq:methods:likelihood}), small spurious ``wiggles'' or concavity flips introduced by sparse sampling or inconsistent point placement can disproportionately affect the pass/fail counts that enter our likelihood.

    Each annotation is converted into an ordered curve parameterized by an affine coordinate $\gamma \in [-1,1]$ that is proportional to the cumulative arc length along the annotation. %
    We then define a $C^2$ continuous spline $\mcal{S}(\gamma; \mbs{\theta})$, where $\mbs{\theta}$ are the knot positions (and two velocity-like terms) needed to specify the track. %
    This spline basis provides a smooth representation that captures the large-scale morphology of the stream. %

    We determine the optimal spline parameters by minimizing a cost function with the desired properties of the track: %
    (1) alignment with the annotated points and %
    (2) suppression of concavity changes. %

    The first contribution to the cost enforces consistency with the annotated points. %
    \begin{equation} \label{eq:shape_cost}
      \chi^2(\boldsymbol{\theta}) = \sum_{i=1}^N \frac{\bigl\lVert \mbf{x}_i - \mcal{S}(\gamma_i;\,\boldsymbol{\theta})\bigr\rVert^2}{\sigma_i^2},
    \end{equation}
    where $\mbf{x}_i$ are the annotation points, $\sigma_i$ are inverse weights specifying how tightly the track should pass through each point, and $\gamma_i$ is the affine coordinate. %
    For a $C^2$ spline the parameters $\boldsymbol{\theta}$ are the knot positions and two $\gamma$-velocity terms.

    The second contribution penalizes rapid changes in the sign of the curvature along the track:
    \begin{equation} \label{eq:concavity_cost}
      \mcal{C}_{\Delta\kappa}(\theta) = \!\int_{-\!1}^{1} \!\! \left( \frac{1}{v}\frac{\rm{d}\widetilde{\mrm{sgn}}(\kappa_\pm)}{d\gamma} \right)^2 \!\! d\gamma.
    \end{equation}
    where $v(\gamma) = |\mbf{T}(\gamma)|$ is the track velocity,
    $\kappa_\pm$ is the signed curvature,
    and $\widetilde{\mrm{sgn}}$ is a differentiable approximation to the sign function,
    implemented as $\widetilde{\mrm{sgn}}(x) = \arctan(\lambda_\kappa x)$ with steepness parameter $\lambda_\kappa$. %
    This term ensures that the curvature is well behaved and avoids concavity changes introduced by human annotation. %

    The full cost function is a weighted sum of these contributions:
    \begin{equation} \label{eq:cost_function}
      \mcal{C}_{\rm tot} = w_{\Delta\kappa} \mcal{C}_{\Delta\kappa} + w_{\mrm{pts}} \chi^2(\boldsymbol{\theta})
    \end{equation}
    with $w_{\Delta\kappa}$ and $w_{\mrm{pts}}$ setting the relative importance of the terms. %
    Minimization of $\mcal{C}_{\rm tot}$ yields an optimized spline $\mcal{S}(\gamma;\boldsymbol{\theta})$ that smoothly traces the stream. %

    The optimized spline serves as the ridge line input to the curvature-based likelihood described next. %
    Further mathematical details of the optimization procedure can be found in \citet{Starkman+:2026:EuclidStreamsPilot} and the method is implemented in \Potamides~ \citep{Potamides:2026}. %

\section{Curvature analysis of remaining host galaxies}\label{app:other_results}

In this appendix, we present the inference results for the remaining systems in our sample (\autoref{fig:data:all_galaxies}).

Based on their distinct morphological characteristics, we classify these galaxies into three categories, (\emph{i}) systems exhibiting multiple tidal features (2MASXJ04470252-4305068 and NGC~1578), (\emph{ii}) systems with single streams that appear visually interrupted by the host galaxy (ESO~287-046, ESO~285-042, NGC~5812, and IC~160), and 
(\emph{iii}) systems exhibiting coherent, quasi-circular loops (2MASXJ12284541-0838329 and NGC~577). We summarize the constraints derived for each group below.

\subsection{Systems with Multiple Tidal Features}\label{app:multi_feature}

\begin{figure*}
\centering
\includegraphics[width=\linewidth]{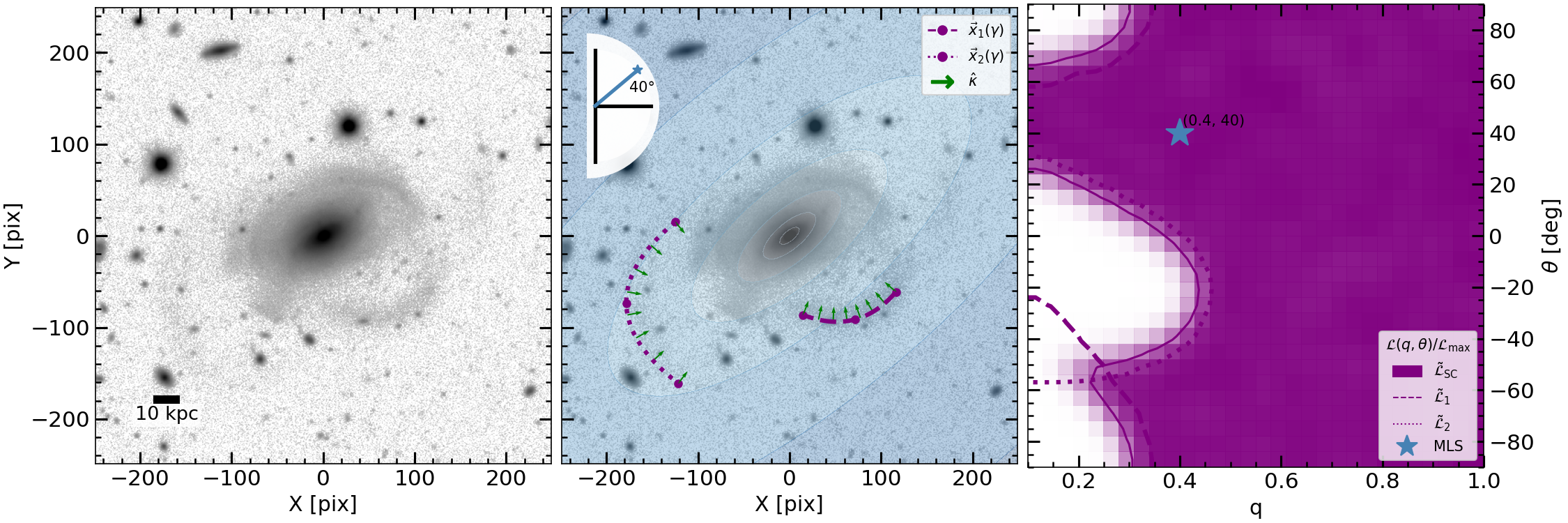}
\caption{
2MASXJ04470252-4305068: Likelihood inference from two stellar streams. %
Colors, symbols, and panel arrangement follow  \autoref{fig:NGC7506}.
In this case, the selected MLS for the halo equipotentials is $(q, \theta)=(0.65,10\degree)$. %
}
\label{fig:2MASXJ04470252-4305068}
\end{figure*}

\begin{figure*}
\centering
\includegraphics[width=\linewidth]{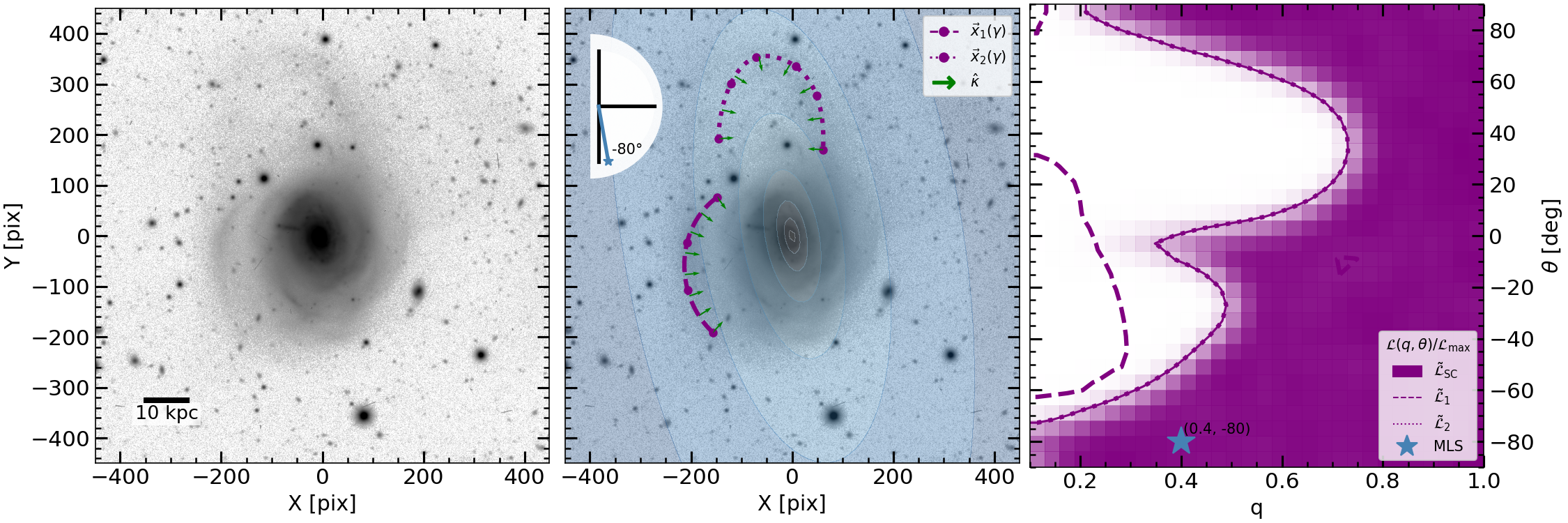}
\caption{
NGC1578: Likelihood inference from two stellar streams. %
Colors, symbols, and panel arrangement follow  \autoref{fig:NGC7506}.
In this case, the selected MLS for the halo equipotentials is $(q, \theta)=(0.4,-80\degree)$. %
}
\label{fig:NGC1578}
\end{figure*}

2MASXJ04470252-4305068 (\autoref{fig:2MASXJ04470252-4305068}) and NGC~1578 (\autoref{fig:NGC1578}) each contain more than one visible tidal feature, which we treat as independent tracers of the same potential. %
Both systems exhibit two distinct features: one resembling a turning-point stream (dotted line), and a ring-like arc (dashed line); the combined constraints are shown as solid contour. 
As expected from \autoref{sec:results:galaxy_weak}, the ring-like arcs provide only weak constraints, ruling out extreme flattenings ($q \lesssim 0.35$) at certain orientations. %
As discussed in \autoref{subsec:sensitivity}, for turning-point streams the constraining power decreases with increasing opening angle. %
The turning-point stream in 2MASXJ04470252-4305068 has a wide opening angle, and can only exclude extreme flattenings ($q \lesssim 0.4$) at certain orientations. %
In contrast, the turning-point stream in NGC~1578 has a much tigher, U-shaped morphology, providing correspondingly tigher constraints that rule out even mild flattening ($q\lesssim 0.7$) at certain orientations. %

\subsection{Visually Interrupted Streams}
\label{appendix:visually_interrupted_streams}

\begin{figure*}
\centering
\includegraphics[width=\linewidth]{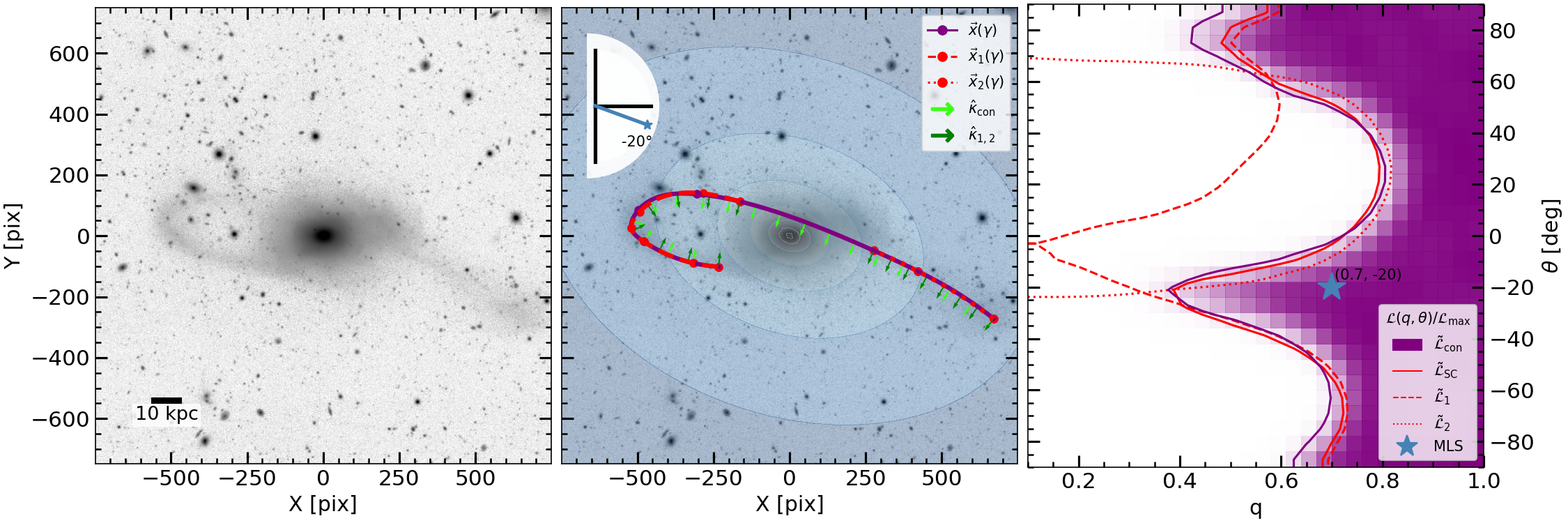}
\caption{
IC160: Likelihood inference from a visually interrupted stream. %
Colors, symbols, and panel arrangement follow \autoref{fig:UGC6397}. %
In this case, the selected MLS for the halo equipotentials is $(q,\theta)=(0.7,-20\degree)$. %
}
\label{fig:IC160}
\end{figure*}

\begin{figure*}
\centering
\includegraphics[width=\linewidth]{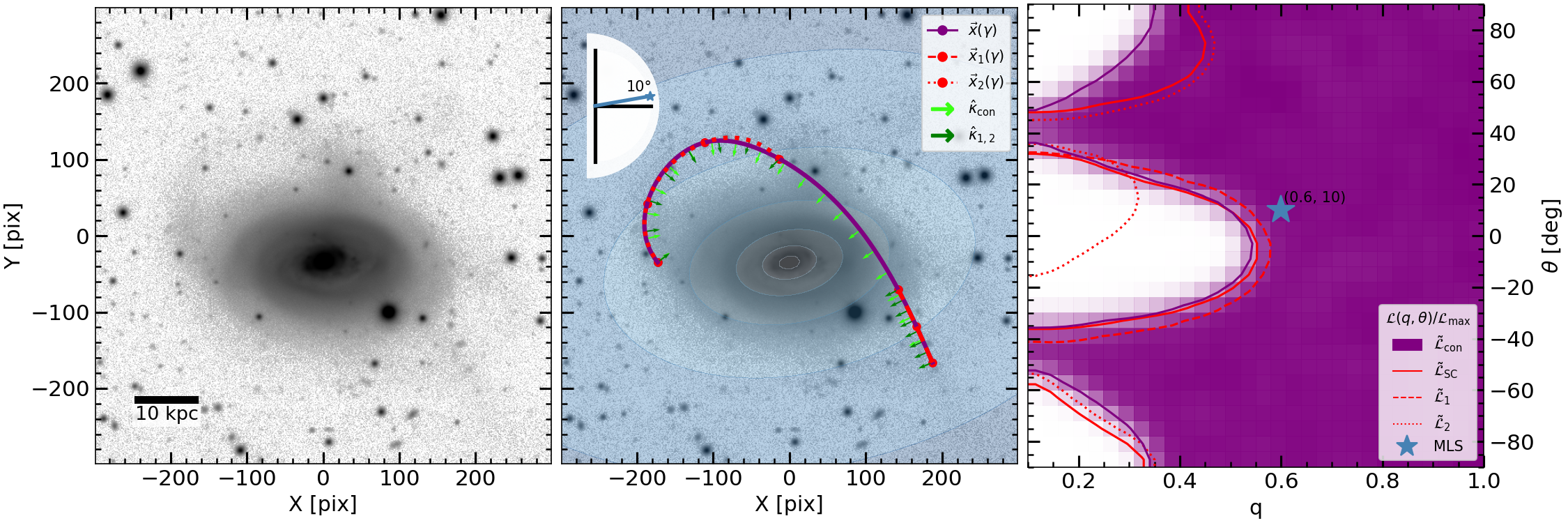}
\caption{
ESO285-042: Likelihood inference from a visually interrupted stream. %
Note: The galaxy center is offset from the origin at $(X,Y)\approx(0,-35)$ pix. %
Colors, symbols, and panel arrangement follow \autoref{fig:UGC6397}. %
In this case, the selected MLS for the halo equipotentials is $(q, \theta)=(0.65,10\degree)$. %
}
\label{fig:ESO285-042}
\end{figure*}

\begin{figure*}
\centering
\includegraphics[width=\linewidth]{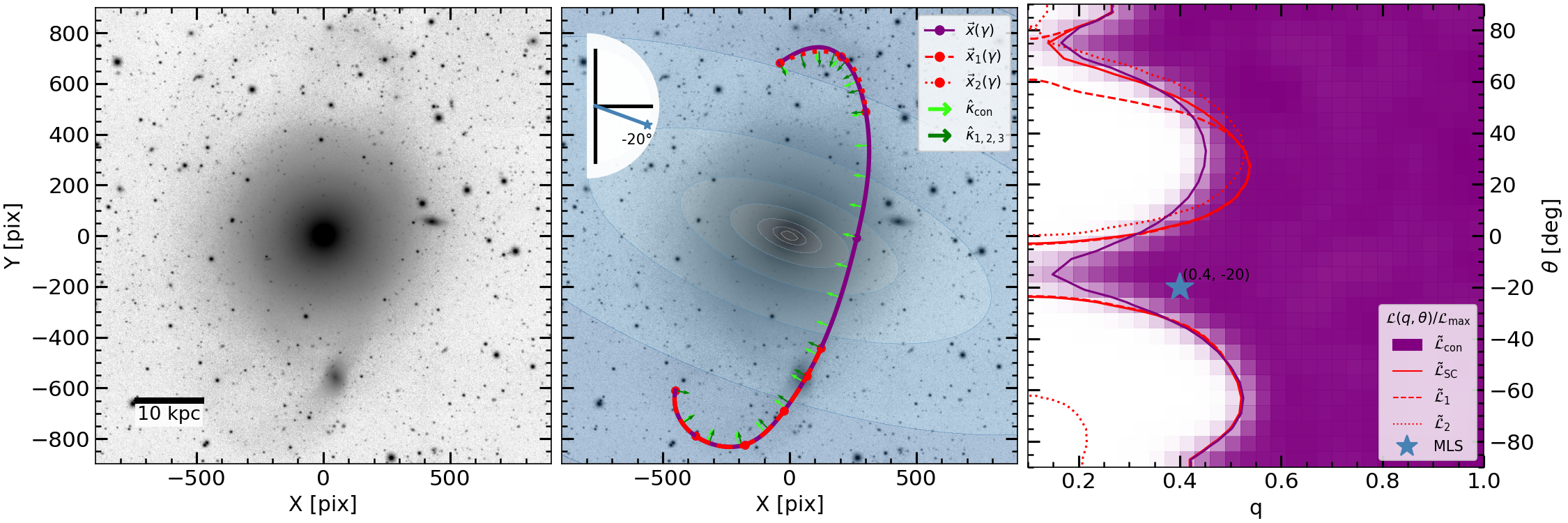}
\caption{
NGC5812: Likelihood inference from a visually interrupted stream. %
Colors, symbols, and panel arrangement follow \autoref{fig:UGC6397}.
In this case, the selected MLS for the halo equipotentials is $(q, \theta)=(0.4,-20\degree)$. %
}
\label{fig:NGC5812}
\end{figure*}

\begin{figure*}
\centering
\includegraphics[width=\linewidth]{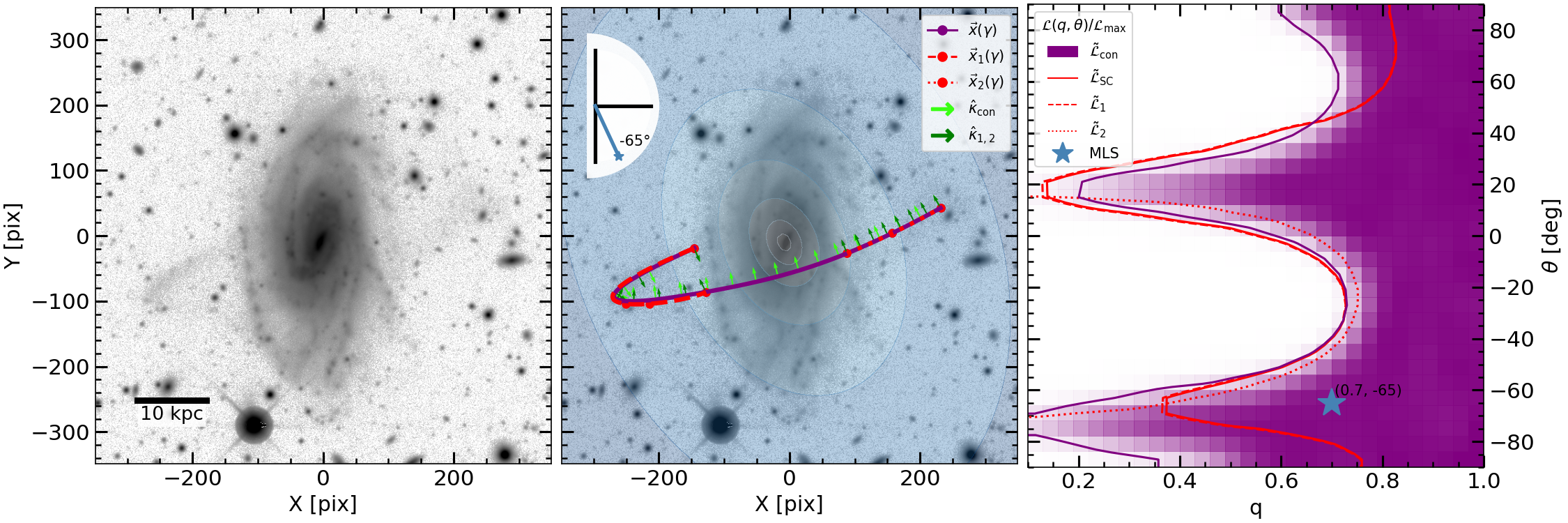}
\caption{
ESO287-046: Likelihood inference from a visually interrupted stream. %
Colors, symbols, and panel arrangement follow \autoref{fig:UGC6397}.
In this case, the selected MLS for the halo equipotential is $(q, \theta)=(0.7,-65\degree)$. %
}
\label{fig:ESO287-046}
\end{figure*}

Four systems in our sample, IC 160  \autoref{fig:IC160}, ESO285-042 (\autoref{fig:ESO285-042}), NGC5812 (\autoref{fig:NGC5812}),
and ESO287-046 (\autoref{fig:ESO287-046}), feature stream tracks that are physically continuous but appear visually broken by the bright stellar light of the host galaxy. Following the methodology established in \autoref{sec:results:galaxy_good}, we present inference results using both reconstruction methods: (\emph{i}) fitting the visible arcs as independent segments ($\mathcal{L}_{\rm SC}$), and (\emph{ii}) fitting a single continuous polynomial bridging the observational gap ($\mathcal{L}_{\rm con}$).

IC160 (\autoref{fig:IC160}),  ESO285-042 (\autoref{fig:ESO285-042}), and NGC5812 (\autoref{fig:NGC5812}) show a high degree of consistency between the segemnted and continuous result, confirming that the inferred constraints are robust to the reconstruction method in these cases. Among them, IC 160 yields the strongest constraints. Its stream traces a long, coherent arc viewed nearly edge-on, a geometry that effectively breaks degeneracies. The likelihood excludes flattenings of $q<0.4$ across all orientations, and rules out $q\lesssim 0.7$ for position angle $\theta \in [0^\circ, 20^\circ]$. In contrast, NGC~5812 and ESO~285-042 exhibit morphologies closer to great circles, resulting in broader likelihood plateaus.
While they do not strongly constrain orientation, both systems reliably exclude extreme flattenings ($q \lesssim 0.4$) across most of the parameter space.

ESO287-046 (\autoref{fig:ESO287-046}) also provides strong constraints overall, excluding $q\lesssim 0.7$ around $\theta\approx -30\degree$. However, unlike the previous systems, it shows a notable divergence between the two methods in the region  $\theta \in [-90^\circ, -60^\circ]$, where the segmented analysis ($\mathcal{L}_{\rm SC}$) yields tighter constraints than the continuous fit ($\mathcal{L}_{\rm con}$). This discrepancy arises from the curvature geometry near the observational gap. In the segmented fit, the curvature vectors at the inner terminus of the left-hand arc point more sharply inward (towards the "turning point"), enhancing the geometric tension with the acceleration field. The continuous fit, by enforcing smoothness across the gap, slightly relaxes this local curvature, resulting in a more permissive likelihood in that specific orientation range.

\subsection{Great circle streams} \label{app:great_circle}

\begin{figure*}
\centering
\includegraphics[width=\linewidth]{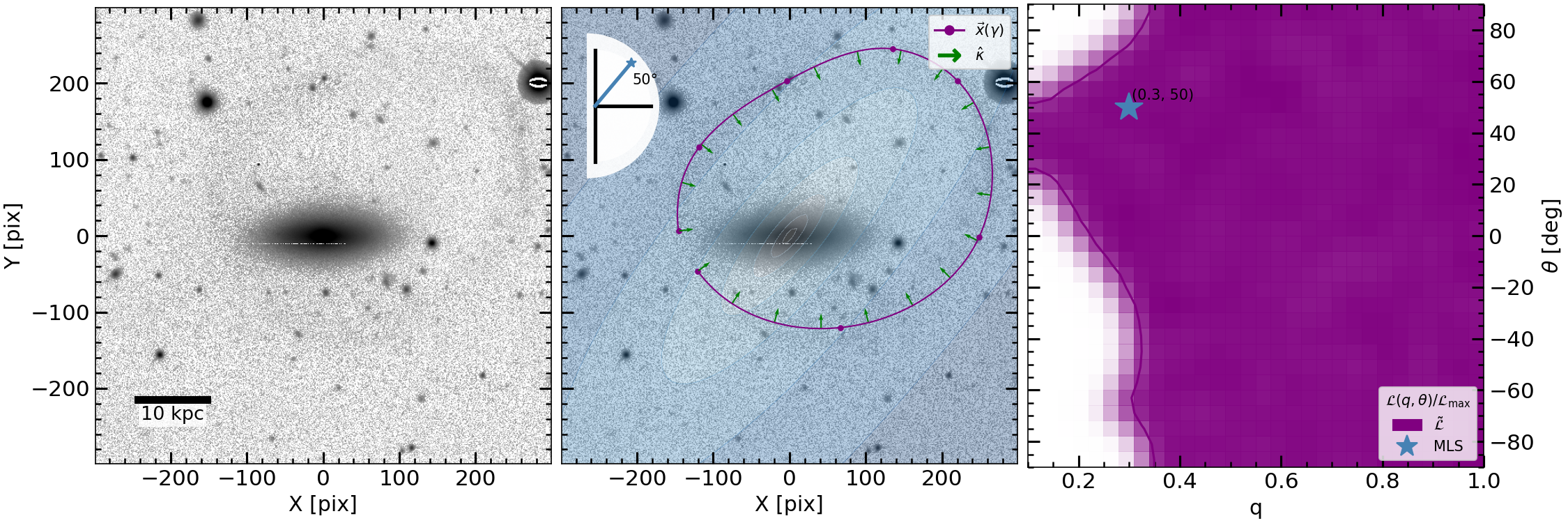}
\caption{
2MASXJ12284541-0838329: Likelihood inference from a single stellar stream. %
Colors, symbols, and panel arrangement follow \autoref{fig:ESO186-063}.
In this case, the selected MLS for the halo equipotentials is $(q, \theta)=(0.7,-50\degree)$. %
}
\label{fig:2MASJ12284541-0837329}
\end{figure*}

\begin{figure*}
\centering
\includegraphics[width=\linewidth]{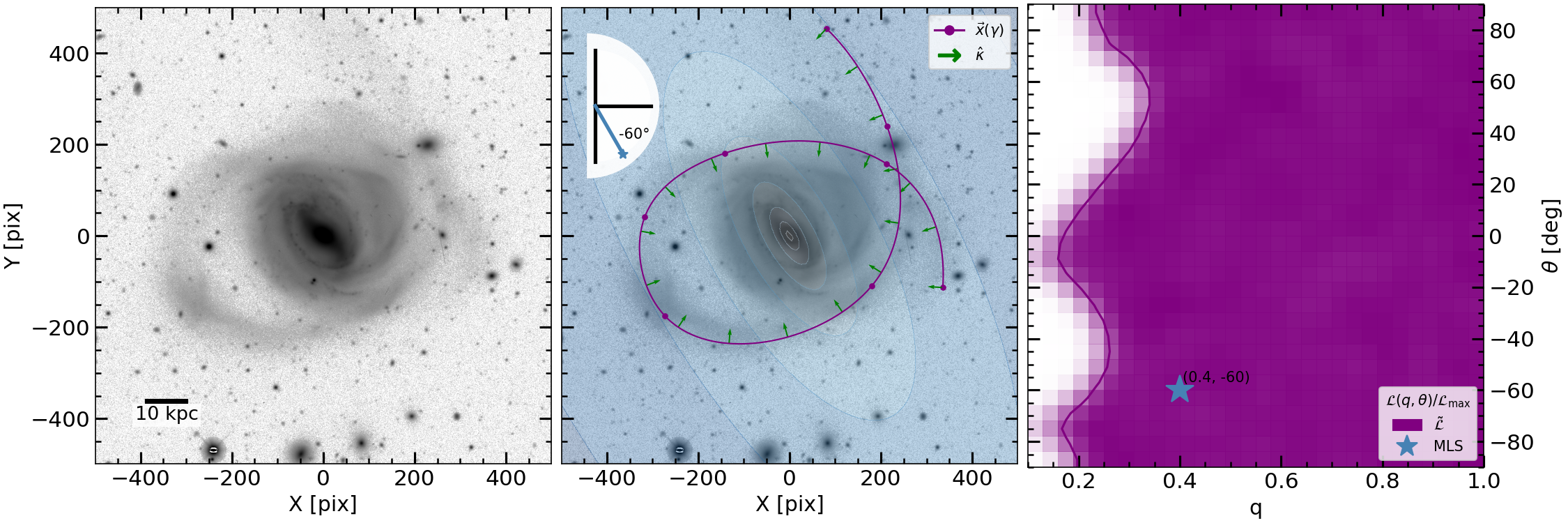}
\caption{
NGC577: Likelihood inference from a single stellar stream.  %
Colors, symbols, and panel arrangement follow \autoref{fig:ESO186-063}. %
In this case, the selected MLS for the halo equipotential is $(q, \theta)=(0.4,-60\degree)$. %
}
\label{fig:NGC577}
\end{figure*}

The streams in 2MASXJ12284541-083829 (\autoref{fig:2MASJ12284541-0837329}) and NGC~577 (\autoref{fig:NGC577}) exhibit ring-like morphology and enclose their host galaxies. %
As discussedin \autoref{sec:results:galaxy_weak}, for these ring-like streams the curvature directions are compatible with the projected acceleration directions over a wide range of halo models, so the method has little sensitivity to the halo geometry. %
Both systems therefore show broad likelihood plateaus spanning nearly all orientations, but they still provide lower limits on the potential flattening, excluding extreme flattenings of $q<0.3$ and $q<0.15$, respectively. %

\section{Geometric Bounds on Intrinsic Shape} \label{app:projection_proof}

    For completeness, in this appendix we derive the geometric constraint that an observed projected axis ratio $q_{\text{proj}}$ serves as a strict upper bound on the intrinsic axis ratio $c/a$. %
    
    Consider a triaxial ellipsoid aligned with its principal axes, characterized by semi-axis lengths $a \ge b \ge c > 0$. %
    This can be represented by a symmetric positive-definite shape tensor $\mathbf{S}$ with eigenvalues $\lambda_1 = a^2,\; \lambda_2 = b^2,\; \lambda_3 = c^2$. %
    When this ellipsoid is orthogonally projected onto an arbitrary plane, the resulting ellipse is described by a $2\times2$ projected shape tensor, $\mathbf{S}_{\text{proj}}$, with eigenvalues $\mu_1 = a_{\text{proj}}^2$ and $\mu_2 = b_{\text{proj}}^2$, where $a_{\text{proj}} \ge b_{\text{proj}}$. %
    
    By the Poincar\'e separation theorem (also known as the Cauchy interlacing theorem; e.g., \citealt{Bellman:1997}), the eigenvalues of any $2\times2$ principal submatrix of a $3\times3$ symmetric matrix interlace with the eigenvalues of the full matrix. %
    Applied to our shape tensor, this yields the chain of inequalities:
    \begin{equation} \label{eq:interlacing}
        a^2 \;\ge\; a_{\text{proj}}^2 \;\ge\; b^2 \;\ge\; b_{\text{proj}}^2 \;\ge\; c^2 \;.
    \end{equation}
    Taking square roots (all quantities are positive) extracts the two bounds needed for our result: %
    \begin{enumerate}
        \item From $a^2 \ge a_{\text{proj}}^2$: the projected major axis satisfies $a_{\text{proj}} \le a$ (projection cannot enlarge the longest axis). %
        \item From $b_{\text{proj}}^2 \ge c^2$: the projected minor axis satisfies $b_{\text{proj}} \ge c$ (projection cannot shrink below the shortest intrinsic axis). %
    \end{enumerate}
    The observed projected axis ratio is defined as $q_{\text{proj}} \equiv b_{\text{proj}} / a_{\text{proj}}$. Applying the two bounds above --- replacing the numerator by its lower bound and the denominator by its upper bound --- we obtain:
    \begin{equation} \label{eq:qproj_bound}
        q_{\text{proj}} \;=\; \frac{b_{\text{proj}}}{a_{\text{proj}}} \;\ge\; \frac{c}{a} \;. %
    \end{equation}
    This shows that the projected axis ratio can never be smaller than the intrinsic short-to-long axis ratio. Physically, projection onto a plane can only make an ellipsoid appear rounder, never flatter, than it intrinsically is. %
    Consequently, measuring a projected flattening $q_{\text{proj}}$ rules out all intrinsic shapes that are rounder than the observation: any model with $c/a > q_{\text{proj}}$ is geometrically prohibited. %

\bibliography{main}{}
\bibliographystyle{aasjournal}

\end{document}